\documentclass[review]{elsarticle}
\journal{Energy}
\usepackage{xcolor,soul,framed} 
\usepackage{lineno,hyperref}
\modulolinenumbers[5]
\colorlet{shadecolor}{yellow}

\DeclareGraphicsExtensions{.pdf,.jpeg,.png}

\usepackage[cmex10]{amsmath}
\usepackage{array}
\usepackage{mdwmath}
\usepackage{mdwtab}
\usepackage{eqparbox}
\usepackage{url}
\usepackage[utf8]{inputenc}
\usepackage{lmodern,textcomp}  

\usepackage{wrapfig}
\usepackage{amssymb}
\usepackage{algorithm}
\usepackage{algpseudocode}

\usepackage{placeins}

\usepackage{longtable}

\newcommand{\red}[1]{#1}

\begin{document}
\begin{frontmatter}
    \title{Implementing Flexible Demand:\\ Real-time Price vs. Market Integration}
\author[cwc]{Florian Kühnlenz\corref{mycorrespondingauthor}}
\cortext[mycorrespondingauthor]{Corresponding author}
\address[cwc]{Centre for Wireless Communications (CWC), University of Oulu, Finland}
\ead{Florian.Kuhnlenz@oulu.fi}
\author[lut]{Pedro H. J. Nardelli}
\address[lut]{Laboratory of Control Engineering and Digital Systems, School of Energy Systems, Lappeenranta University of Technology, Finland}
\author[obs]{Santtu Karhinen}
\author[obs]{Rauli Svento}
\address[obs]{Business School at University of Oulu, Finland}
%
%

\begin{abstract}

This paper proposes an agent-based model that combines both spot and balancing electricity markets. From this model, we develop a multi-agent simulation to study the integration of the consumers' flexibility into the system. Our study identifies the conditions that real-time prices may lead to higher electricity costs, which in turn contradicts the usual claim that such a pricing scheme reduces cost. 

We show that such undesirable behavior is in fact systemic. Due to the existing structure of the wholesale market, the predicted demand that is used in the formation of the price is never realized since the flexible users will change their demand according to such established price.

As the demand is never correctly predicted, the volume traded through the balancing markets increases, leading to higher overall costs. In this case, the system can sustain, and even benefit from, a small number of flexible users, but this solution can never upscale without increasing the total costs. 

To avoid this problem, we implement the so-called "exclusive groups". Our results illustrate the importance of rethinking the current practices so that flexibility can be successfully integrated considering scenarios with and without intermittent renewable sources.
\end{abstract}

\begin{keyword}
real-time pricing\sep day-ahead market\sep balancing market\sep market design\sep demand response\sep smart grid
\end{keyword}

\end{frontmatter}
\section{Introduction}

In our free market economies, monetary incentives are usually seen as a way to change human behavior. 
This assumption is widespread, going from cost-benefits decisions of our everyday life to economic policies of countries. 
It seems natural, hence, to assume that exposing end-users of electric power to varying prices will inevitably result in behaviors that maximize the efficiency of the electricity system.
One of the earliest mentions about this idea can be found in \cite{Houthakker:1951fv} and \cite{Steiner:1957vo}.
Both articles discuss the problem of peak loads and pricing schemes that discourage customers from using electric power during such high demand periods.
The proposed two-tariff system has been implemented afterwards and is still in use.
For example in the UK, the so-called ``Economy 7'' tariffs  have a cheaper electricity price for seven hours during the night \cite{Hamidi:2009ck}.
There, such tariffs are typically used in conjunction with storage heaters to store cheap power from the night to heat the house during the day.

In recent years this idea has been taken further: the customers in some places have the option to choose the so-called real-time price \cite{savolainen2012real}.
As the name suggests this price should reflect the cost of producing electricity in ``real-time'' and therefore allow the customers to allocate their flexible demand (e.g. washing machines, dish washers or charging of electric cars) during times where power is cheaper. 
This would then align the needs of the system, e.g. reducing peak load to avoid overloading of power lines, with the interest of consumers to save money, assuming of course that the price correctly reflects any bottlenecks in production or transmission \cite{Houthakker:1951fv}.
The concept of a real-time price is also in the core of many concepts of the evolving power grid, the smart grid, where decentralization and intermittent production drive the need for new coordination mechanisms \cite{ELMA2017206}.
This development however seems curious when compared to the findings of \cite{Houthakker:1951fv} back in 1951: ``We shall not proceed to consider tariffs with three or more running charges, as the introduction of further rates yields rapidly diminishing returns, while consumer costs increase accordingly.''

From the current technological perspective in combination with the liberalized electricity system, this argument however looks inappropriate: not only \red{does} the market generate public price information with at least hourly resolution, but also there exists technological means to inform and charge the customer with the same time resolution.
Therefore, the idea to expose customers to these hour-by-hour prices seems straightforward.

Yet, the power system is changing in two fundamental ways with the introduction of renewable energy sources.
Power is no longer produced only by centralized stations, but rather it is becoming more and more collected in \red{a} distributed manner due to the increasing number of intermittent sources \red{like} solar and wind \cite{Hirth:2013td,Hirth:2013gca}.
Hence, the power system needs to coordinate a fast growing number of players, while the volatile nature of renewable sources requires systemic changes to be properly accommodated.
This fact implies that models that assume a production that creates constant merit order curves cannot be applied nowadays.
For example, the work done by \cite{Borenstein:2005et} in fact provides a sound model for the time and place it was developed, but is in its basic form unfit \red{for} the reality of the modern power grid in liberalized markets with increasing number of renewable sources (extensions have been proposed for example in \cite{kopsakangas2013economic,kopsakangas2013promotion}.

The price in the Nordic power system is known to be influenced by hydro power as shown in \cite{Botterud:2010dd} based on eleven years of historic data. 
The hydro reservoirs in Sweden, Norway and Finland behave like traditional power plants, so that production can be scheduled to follow the power demand\cite{huuki2017flexible}. 
Germany, in its turn, has seen negative prices for power in the recent years \cite{Nicolosi:2010vt} due to the big amounts of wind and solar that have been added to the system.
These prices occur since wind and solar cannot be scheduled in the traditional way; the power output can only be capped.

The challenge then becomes to use or store the renewable power whenever it is available.
This would bring many positive effects from both environmental and economical perspectives: low price electricity (almost zero-marginal cost source) with low CO2 emission levels.
This architectural change, however, should not be underestimated.
In a system dominated by traditional production, a peak in consumption shall be avoided in order to reduce the investment in the grid capacity and the usage of peak power plants with high production costs.
In a system dominated by variable renewable sources, conversely, a peak in consumption may be the way to optimally utilize the available power (as in a case of strong winds and their respective power outputs).

At the first sight, this is a good argument in favor of exposing consumers to the variable prices of the electricity market: they are supposed to correctly reflect the availability of renewable power or the use of expensive power plants in peak periods.
The real-time price would then set the right incentives for consumers to allocate their flexible usage or storage capabilities, creating a win-win situation for the consumer and the grid.

This way of reasoning is notwithstanding flawed in the sense that it misconceives the day-ahead price as a reflex of the realized demand, instead of the predicted demand.
Flexible consumers guided by the real-time (i.e. spot) price realize their consumption differently from the prediction that in fact defined such a price.
Consequently, predictions may not be realized and balancing becomes more necessary, which may result in higher prices as the electricity is more expensive in balancing markets.

This work demonstrates how the current (incorrect) approach to real-time pricing (specially for retail customers) cannot as such be sustained when the ratio of flexible consumers grows, while the incentives are incorrectly aligned with the model pursued in many EU countries.
As an effective solution to this systemic failure, we show that flexibility (in large scale) can be successfully incorporated into the system by daily profile bids instead of independent hourly bids.
Such a solution is in fact viable and already implemented in liberalized markets as the so-called ``smart blocks'' (used in EPEX SPOT \cite{EPEXSPOTIntroduct:2014tm}) or ``exclusive groups'' (used in Nord Pool \cite{Dourbois:2015fd,nordpool:blc}). 
They work by informing the market about different possible demand profiles, that can be realized due to the available flexibility.
By covering these issues, the present work contributes to a better understanding of how flexibility shall be integrated into the market considering scenarios with and without intermittent renewable sources.
\red{Further this work introduces a reusable open source model that provides unique capabilities of modeling production (renewable and conventional) and load on a minute by minute basis.
It can be also used to analyze the power system in a holistic way by modeling not only the day-ahead market but also the balancing market, therefore respecting the physical constraints of matching generation and consumption at all times.}

The rest of this article is divided as follows.
In Section \ref{sec-model}, we describe the proposed agent-based model incorporating both spot and balancing markets, which is used to carry out the present study.
Section \ref{sec-results} presents the numerical results supporting our claims, while Section \ref{sec-disc} discusses the implications and limitations of this work.
Section \ref{sec-concl} concludes this paper.

\section{Model}
\label{sec-model}

In the electricity grid, supply and demand need to be balanced due to physical constraints \cite{nardelli2014models}. 
To ensure this match in a liberalized market, electricity is traded in several stages with increasing time resolution. 
Trading happens in the following stages \cite{nordpool:blc}: day-ahead market, intra-day market and balancing market. 
All differences that could not be accounted for in the day-ahead market need to be corrected. 
This may occur in the intra-day market, but the actual realized imbalances must be corrected at the balancing market, in real time (minute-scale in contrast to hour-scale from spot market). 

The structure of this market system has been developing over time and reflects not only the idea of a free and efficient market with competition (e.g. \cite{Nicolaisen:2001hy}) but also the engineering needs of the power system. 
The different time resolutions and lead-times allow for proper long-term planning of production for the types of power plants that are less flexible, while close to real-time markets allow the for final adjustments driven by the physical reality of the situation.
We base our analysis on an agent-based model which was firstly introduced in \cite{kuhnlenz2017agent} and will be extended with several features explained below.
It includes producers, utilities and consumers, while evolving around three stages of interaction between them. 
These interactions are indeed the reason why we choose an agent-based model for our analysis instead of a more traditional analytic model.
Agent-based models \cite{Helbing:2011th} have been used to analyze power systems before as in \cite{Kremers:2013vr, Kok:2013uo, Santos:2012hv, Gallo:2016ev, Santos:2015ir} and have also provided a deeper understanding of the behavior of markets \cite{Helbing:2010bw, LeBaron:2016ht, Farmer:2009cd, lebaron2006agent}. 

We see our model aligned with the previous research in the topic, but with some distinctive differences.
First, we aim to include the physical realities of power systems into the market simulation by using a minute timescale instead of a hourly based one, allowing for the inclusion of the balancing market that covers all deviations from the day-ahead schedule. 
We deem this important as fluctuations of renewable power (e.g. \cite{Milan:2013hg,Anvari:2016bj}) and consumer behavior (e.g. \cite{mclaughlin2011protecting}) work on a sub-hourly timescale even down to a second or sub-second level. 
While fluctuations on the sub-minute level will mostly be covered by primary regulation (therefore  by automatic systems and physical effects in the power system), they can be neglected for a market simulation (although some mechanisms of primary regulations like automatic and manual Frequency Restoration Reserves can function as a market where the Transmission System Operators collect offers from suppliers and use only the cheapest ones).
We decided only to handle the mechanisms that work upwards from a reaction time of a minute as part of the balancing market.
From the perspective of the present work, this still preserves the physical reality that the system needs to be in balance while also reflecting the market-based procurement and dispatching of these capacities.

Second, we see our model not as a tool to forecast actual prices but rather to understand systematic effects and relations.
The agent-based modeling approach is suitable with that premise as it allows not only to capture the interactions between all players in a given system, but it also allows these interactions to have non-optimal and unstable results.
It is also unnecessary to assume that prices or installed capacities are according to the long-term equilibrium,  but they will be shaped to reflect the local reality of any given agent.

In its current form, the model excludes the intra-day market stage and financial markets.
While intra-day markets have been shown to be more important with a higher penetration of renewable sources \cite{Ocker:2015wg}, they are unnecessary to understand the general (qualitative) patterns inside the cascade market system.
This aspect shall be introduced in a further extension of the present model.

Financial markets play a different role in different systems.
In the Nord Pool area, they can only be used as a price hedging instrument but not for scheduling physical flows\footnote{http://www.nordpoolspot.com/the-power-market/Financial-market/}. 
Consequently, they might have an influence on the bidding behavior of market players, but less on the interaction between different markets.

Finally, the goal of our modeling is also to follow the actual implementation of the Nord Pool market (as well as all other European electricity markets) as close as possible.
Our approach tries in some sense to fill the important gap indicated in \cite{Ringler:2016jt}: ``Given the narrow focuses, interactions between different layers, e.g. with centralized wholesale electricity markets, are sometimes neglected. Thereby, a consistent and comprehensive evaluation of effects from smart grid structures is still difficult. Similarly, the concrete link to current market designs and implementation steps are missing, which is why some research papers remain rather academic. Issues of local acceptance are neither addressed explicitly.''

\subsection{Interaction}

\begin{figure*}
\includegraphics[width=0.495\columnwidth]{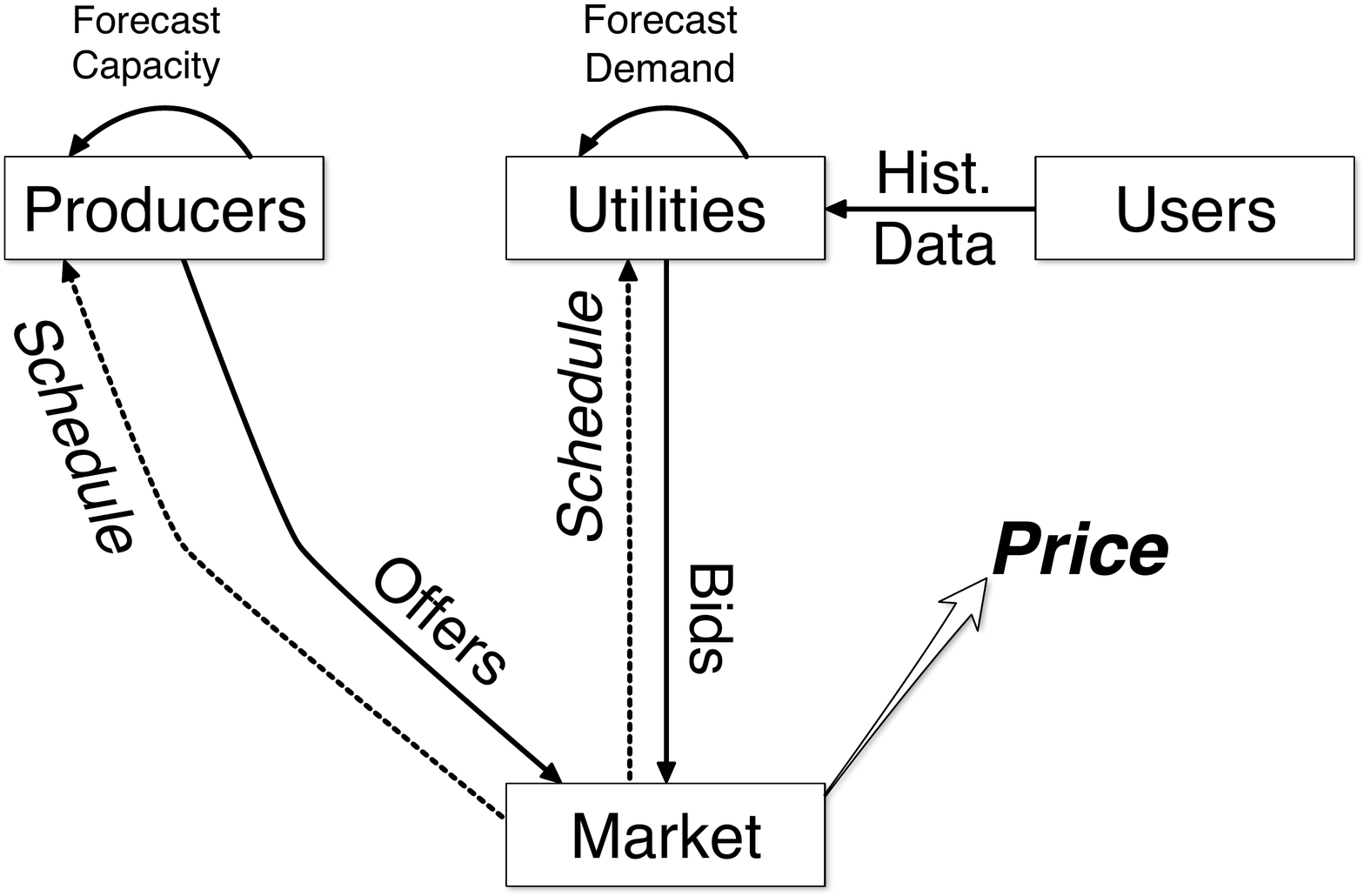}
\includegraphics[width=0.495\columnwidth]{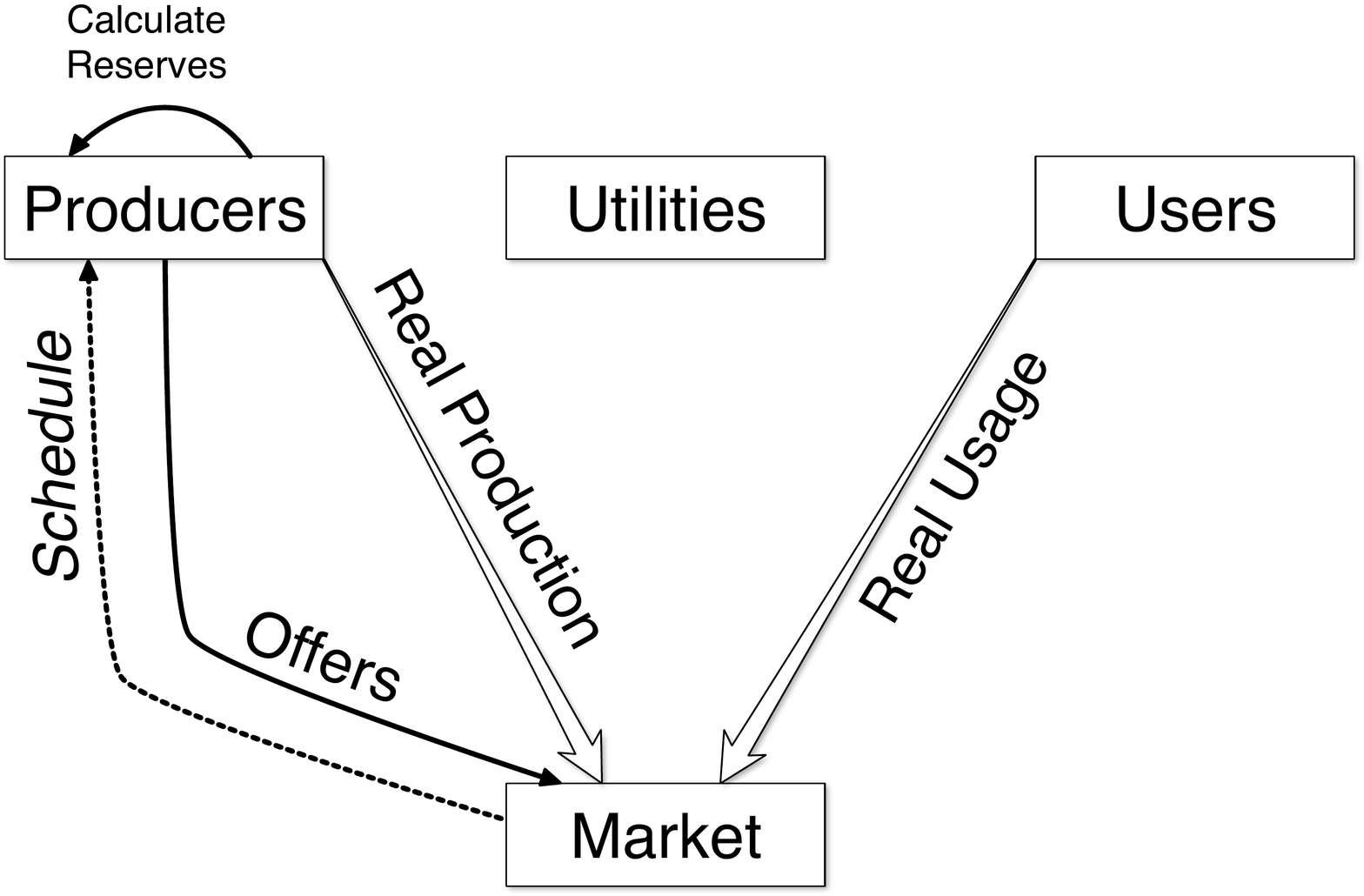}
\centering
\includegraphics[width=0.495\columnwidth]{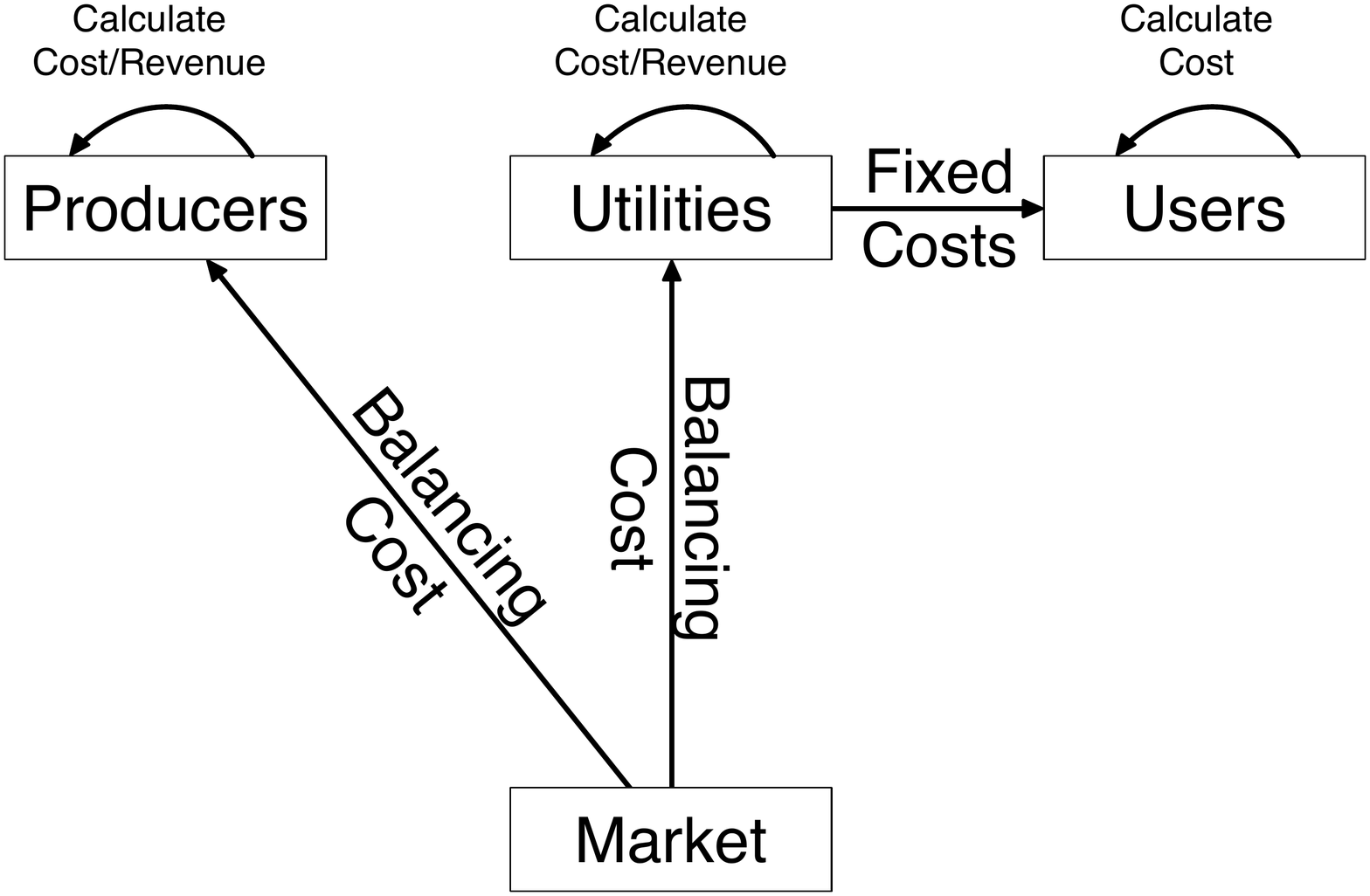}
\caption{\label{interaction}The three interaction stages of the modeled market system. Top left: the day-ahead market period where demand forecasts are matched against production forecasts. Top right: the balancing market period where all deviations are adjusted for. Bottom: the monetary settlement period.}

\end{figure*}

Our model is based on three stages of interaction depicted in Fig. \ref{interaction}.
At first, the utilities make a forecast about their users' consumption to be then submitted to the market as a bid.
Forecasts are based on a historical average with an exponential weight factor so that more recent history is more important for the forecast.
The day-ahead market matches bids from these forecasts with the offers from producers, creating a hourly production schedule for the next day.
This process creates a price (the spot price) for each hour as the most expensive power plant that needs to run during any given hour \cite{Chatzigiannis:2016hz}.
Since all producers bid in their marginal cost of production this price also reflects the production price.

Next, all mismatches between realized and forecasted production and consumption are calculated for every fifteen (15) minutes in the model.
If the mismatch exceeds a certain value it will be balanced by a matching offer from a producer at the balancing market. 
Every producer can offer power into the balancing market, given certain internal limitations, to reflect the behavior of different types of power plants.

All producers have a minimum run requirement that determines if they can offer balancing energy even if they were not dispatched for that hour in the day-ahead market.
They also have a regulation factor that determines how much they can individually change their output for balancing from the value dispatched during the day-ahead period.
Furthermore, every producer will bid its marginal cost with an additional factor into the balancing market.
Producers can also submit several offers for the same period, for example with a certain amount of MWh for a lower and an additional amount for a higher price.

In the final stage, all settlements are accounted for monetarily.
This means all transactions according to the market are executed, so producers get payed for their production while utilities pay for their consumption.
If any party deviated from its day-ahead schedule, it has to pay the balancing price, accordingly.
For this purpose, the Nordic model consists of a two-price system for producers and one-price system for consumers \cite{Vandezande:2010cz}.
The users are all assumed to be under a real-time pricing regime, so rates are according to the day-ahead prices (which is a common practice in the Nordic area).
Users are charged according to their actual consumption of any given day.
The balancing costs will be paid by the utilities, as the market participants, for any deviations from the agreed schedule. 
However, they will be subsequently charged from the users as fixed costs, equally shared between all users of a given utility.

\subsection{Flexibility}

\begin{figure*}
\centering
  \includegraphics[width=0.7\textwidth]{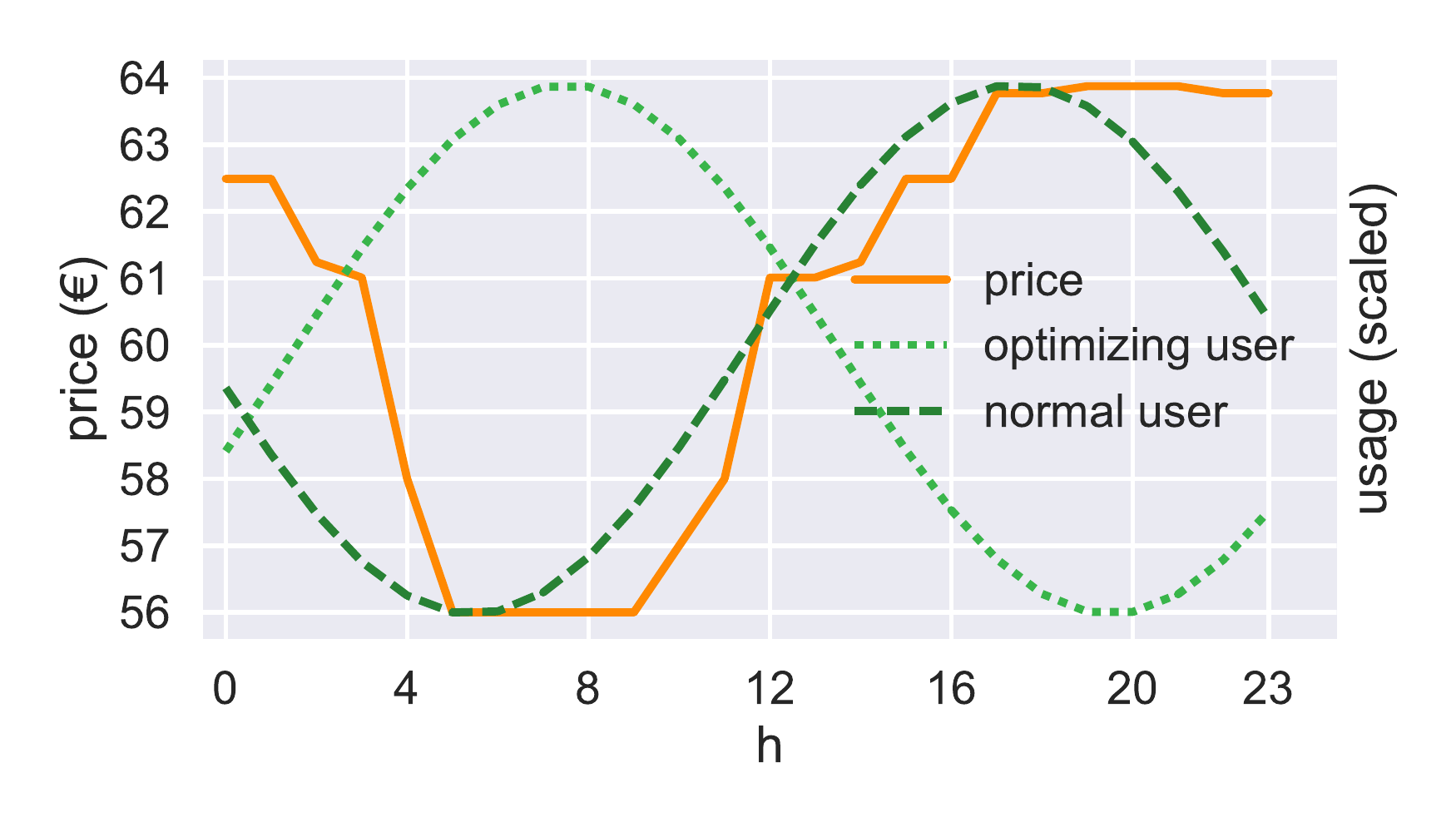}\\
 \caption{\label{user_sine}Load curve examples of an optimizing, flexible user and an ordinary user in the RTP regime. The optimizing user shifts his load by adjusting the phase of the sine curve to have minimum correlation with the price curve. Note: the usage is scaled and not zero based.}  
\end{figure*}

Different from our previous work \cite{kuhnlenz2017agent}, this paper considers that users have the ability to change their electricity usage.
This flexibility can be utilized to shift power consumption away from high-price times to low-price ones to minimize consumer costs. 
As a simplification, the usage patterns are modeled as sine curves so that optimizing agents need to set the phase of their respective sine curve to simulate a shift in the usage pattern, illustrated in Fig. \ref{user_sine}. 
This does not change the shape of the use, but rather simulates the shift of peak power to a different time of the day. 
This is also consistent with the rebound effect \cite{gelazanskas2014demand}, which considers that a reduction of the load during one part of the day is usually followed or preceded by a higher consumption to compensate for the reduced usage.
\red{While the assumption of simple sinusoidal load curves might seem too simplistic considering highly sophisticated models for loads and their flexibility like \cite{Zhong:2015ku,Zhong:2015hj,Srikantha:2012gi}, our aim is to present a neglected systemic effect without any unnecessary complications. In others words, we have targeted to build a model that is ``as simple as possible but no simpler.'' This model would serve as basis for further studies, these involving more realistic models. Nevertheless, to test the consistence of the proposed scenario, we provide preliminary results of a more intricate way of modeling load curves  in the appendix \ref{app-model}, which shows a similar behavior to the ones presented along our main results.} 

To find the optimal shift time, each consumer calculates a correlation of its own usage and the hourly prices of the day-ahead market at the beginning of each simulation day after the prices of the day-ahead market have been announced. 
The position of the minimum value in the resulting correlation vector then marks the optimal shift time, as usage during low price times are maximized and minimized during high price times. 
\red{We also assume that the flexible load is completely automated. Due to the short-term changes of power availability from renewable sources, only automated reactions can provide the needed reaction time. We will analyze this assumption further in the discussion part of this article.}

\subsection{Exclusive Groups}

\begin{figure*}
\centering
  \includegraphics[width=0.8\textwidth]{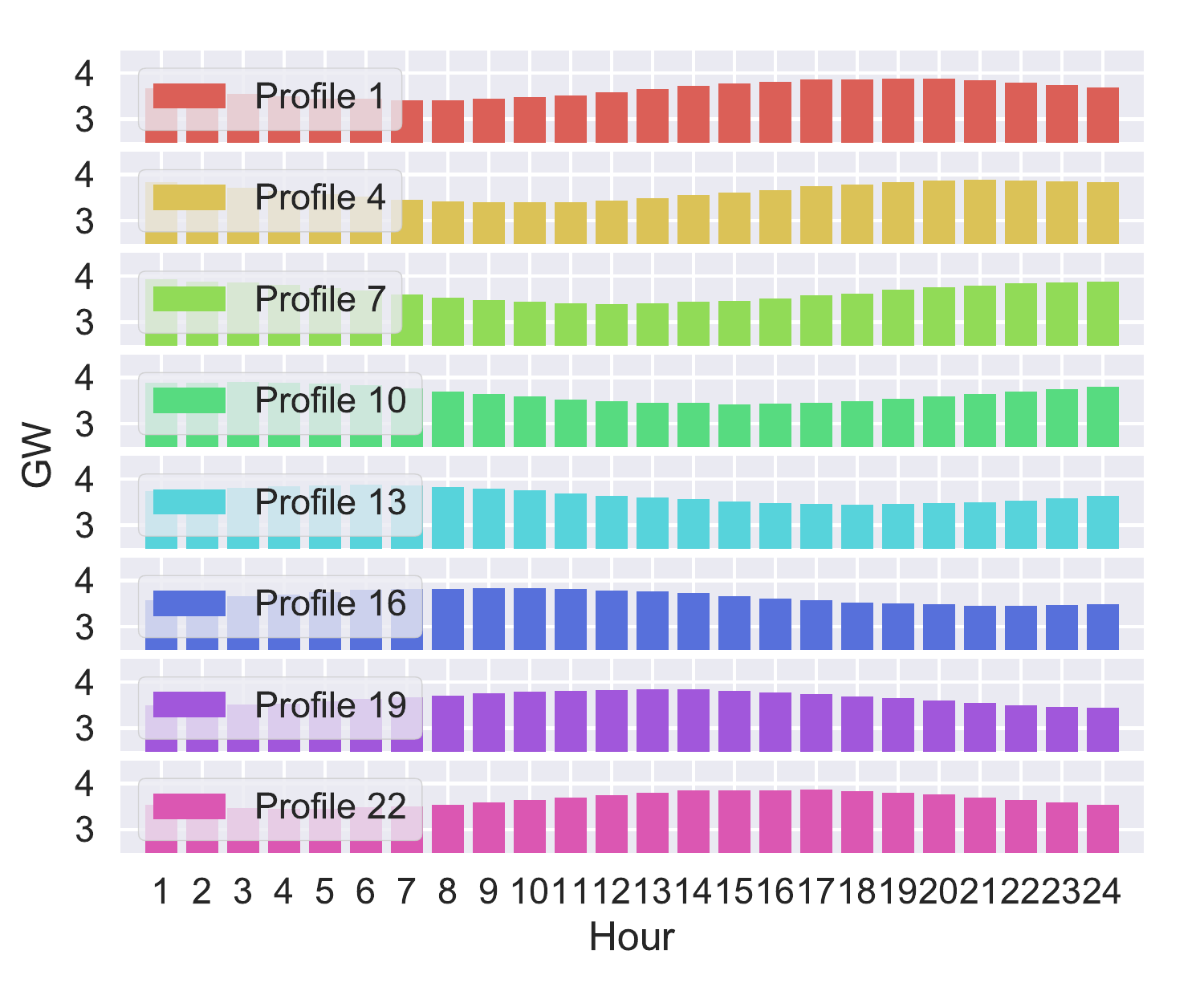}\\
 \caption{\label{profiles}Different profiles of a single utility. The market algorithm has to select the profile which will deliver globally optimal results.}  
\end{figure*}

As to be discussed in the next section, the user-centric, price-based, optimization may lead to undesirable (and usually unexpected) outcomes. 
As we shall demonstrate, these outcomes could be mitigated by the use of so-called ``exclusive groups'' (EXG) \cite{euphemia-PCR} (or ``smart blocks'').
EXGs are one possible way (together with flexible orders or linked block orders \cite{euphemia-PCR}, which are not implemented here)  of biding flexibility directly into the day-ahead market by offering several different groups of hourly offers/bids.
The market matching algorithm then selects exactly one of these profiles, which the agent then has to follow.
An example of an EXG bid  (formed by several profiles)  is shown  in Fig. \ref{profiles}.
In this way, flexibility is directly visible for the market, which shall select the set of profiles that optimize social welfare. 
The details of implemented calculation can be found in the model\cite{sourceCode} and as pseudo-code in  Appendix \ref{model-description}.  

It is worth stressing that this type of offer is not an idea of the authors, but rather an already available product in several European day-ahead market places, \red{yet almost ignored in research.}.
Public documentation shows its intended usage \cite{EPEXSPOTIntroduct:2014tm} and also indicates a growth in its use \cite{PCRStatusUpdate:2016tl}.
Note also that this solution presupposes that the flexibility can be fully used by the utilities. 
In other words, the utilities shall have the control of flexible loads in order to realize all  proposed curves.
Utilities may implement this by direct load control, by internal pricing strategies, or by some color flagging scheme. 
Although this is of core importance for the actual implementation and upscaling of this solution, we assume here that the utilities can deliver the bidding curves and their users are charged following the market output.

\subsection{Model Setup}

\begin{figure*}
\centering
  \includegraphics[width=0.9\textwidth]{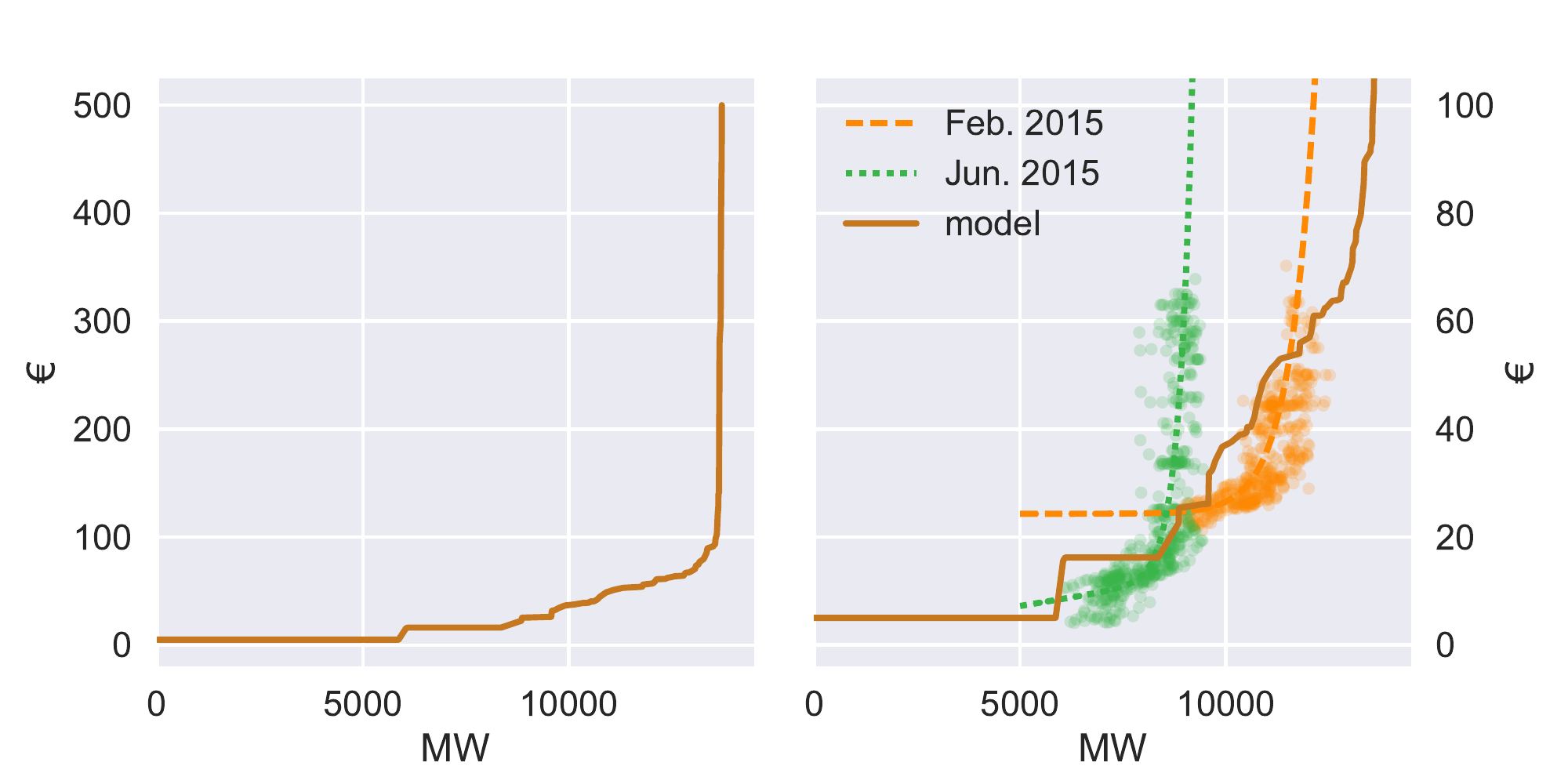}\\
 \caption{\label{merit_order}Left: the merit order curve used for the day ahead market. Right: part of the day ahead merit order curve compared to 2015 data from Nord Pool for the Finnish day ahead market including a fit though the data.}
\end{figure*}

As mentioned before, we do not intend to forecast price, but rather contribute to the understanding of systematic effects when integrating flexibility in the electricity market.
To do so, we shall set the proposed model to qualitatively behave in a sound manner, while not necessarily providing quantitatively realistic results.
Therefore, although we are not targeting at a precise price forecast, we do need to calibrate the proposed multi-agent simulation with suitable data.

The core of the setup is the merit order curve that defines prices at the day-ahead market and, to a certain extent, at the balancing market.
%
The used merit order curve, as well as a comparison to real-world data taken from Nord Pool, can be seen in Fig. \ref{merit_order}. 
The comparison to Nord Pool prices is not only there to show that the general shape of the merit order curve is valid, but also to discuss another interesting effect.
As can be seen by the data from June, peak time prices during summer can be as high as during winter, even though the average consumption and prices are much lower. 
The reasons for this effect are possibly several: 
\begin{itemize}
\item the power plants that run during peak consumption hours are the same during summer and winter,
\item hydro power in the Nordic system is less used during the summer,
\item co-generation is less used during summer,
\item producers might adjust their prices predicting the peak usage.
\end{itemize}
Whatever the case, prices during peak consumption hours can be high even if the overall consumption falls.

For the balancing part, we also included high price power producers at the top end of the merit order curve (not included in Fig. \ref{merit_order}). 
This serves for two purposes.
First, they incorporate the reserves of the TSO procures which are typically used only once the market balancing power is used up.
Second, they represent a kind of buffer that the system will run into when a large amount of balancing power is needed. 
This power then comes at a (very) high price, as it would be in the real world \cite{Fingrid:2016us} (up to the upper price limit of Nord Pool 3000 €/MWh).

In the case of a lack of available up-balancing energy, the model is set up to stop the simulation with an error, while in the real world this would typically call for more extreme measures like load-shedding or even a black-out.
However, such kind of black-outs almost never occur in the real world.
We therefore added a big (1 GW) and expensive (3000€/MWh) power plant at the top of the balancing merit-order curve to represent the aforementioned problematic situation (so our simulation will never reach a black-out or brown-out, but this ``extreme case'' will be captured by a very expensive power).

On the consumer side, we setup the system with a peak to peak difference of about 14\%, which is set based on the maximum to minimum difference in daily load-cycle in the Finnish power system in the data set from 2015 that we analyzed.
The maximum consumption is 12.6 GW.

\section{Results}
\label{sec-results}
\begin{figure*}
  
  \includegraphics[width=\columnwidth]{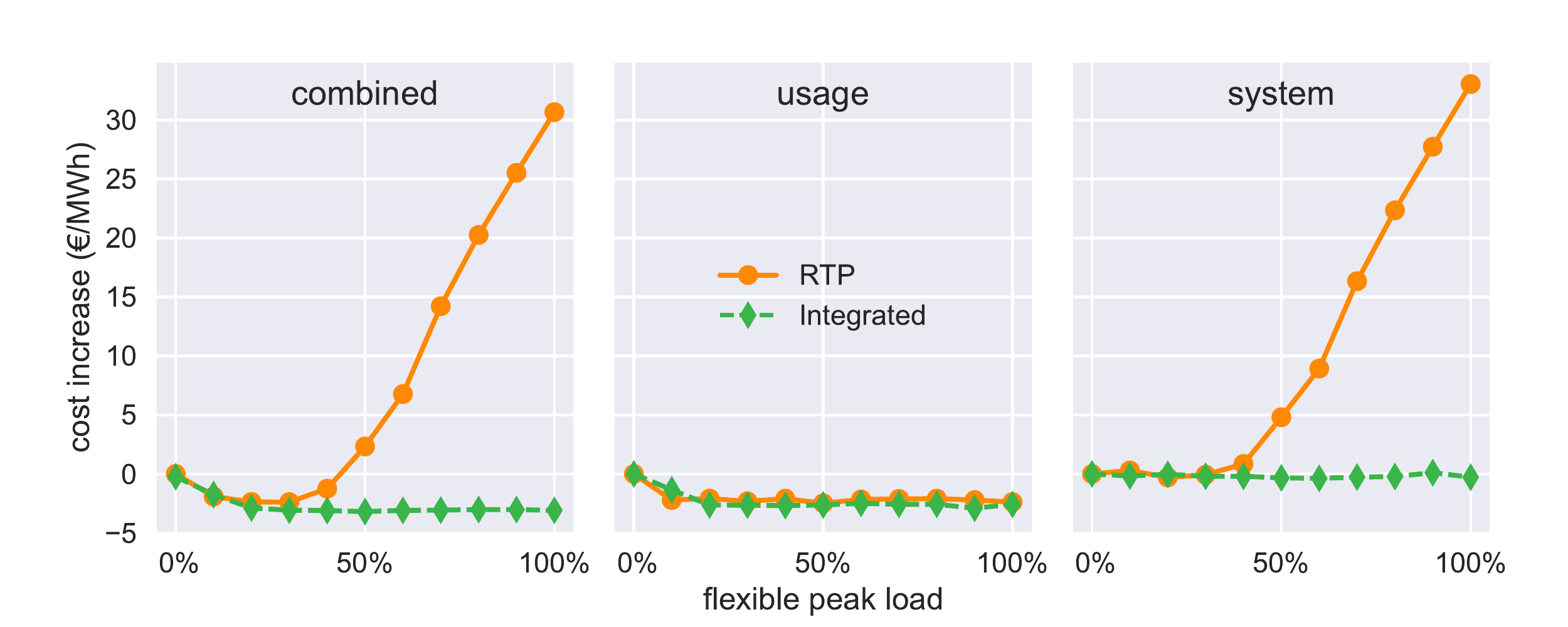}\\
 \caption{\label{savings_split}Price changes of an average MWh with a rising share of flexible peak load in the system. 
Fist figure in the left displays combined costs per MWh resulting from usage costs based on day-ahead prices and the shared balancing costs. 
Central plot shows only the development of usage costs, which result purely from the tariff, which in turn is based on the day-ahead prices. 
Right figure shows shared costs resulting from balancing needs and forecast errors.}

\end{figure*}

\begin{figure*}
\centering
  \includegraphics[width=\textwidth]{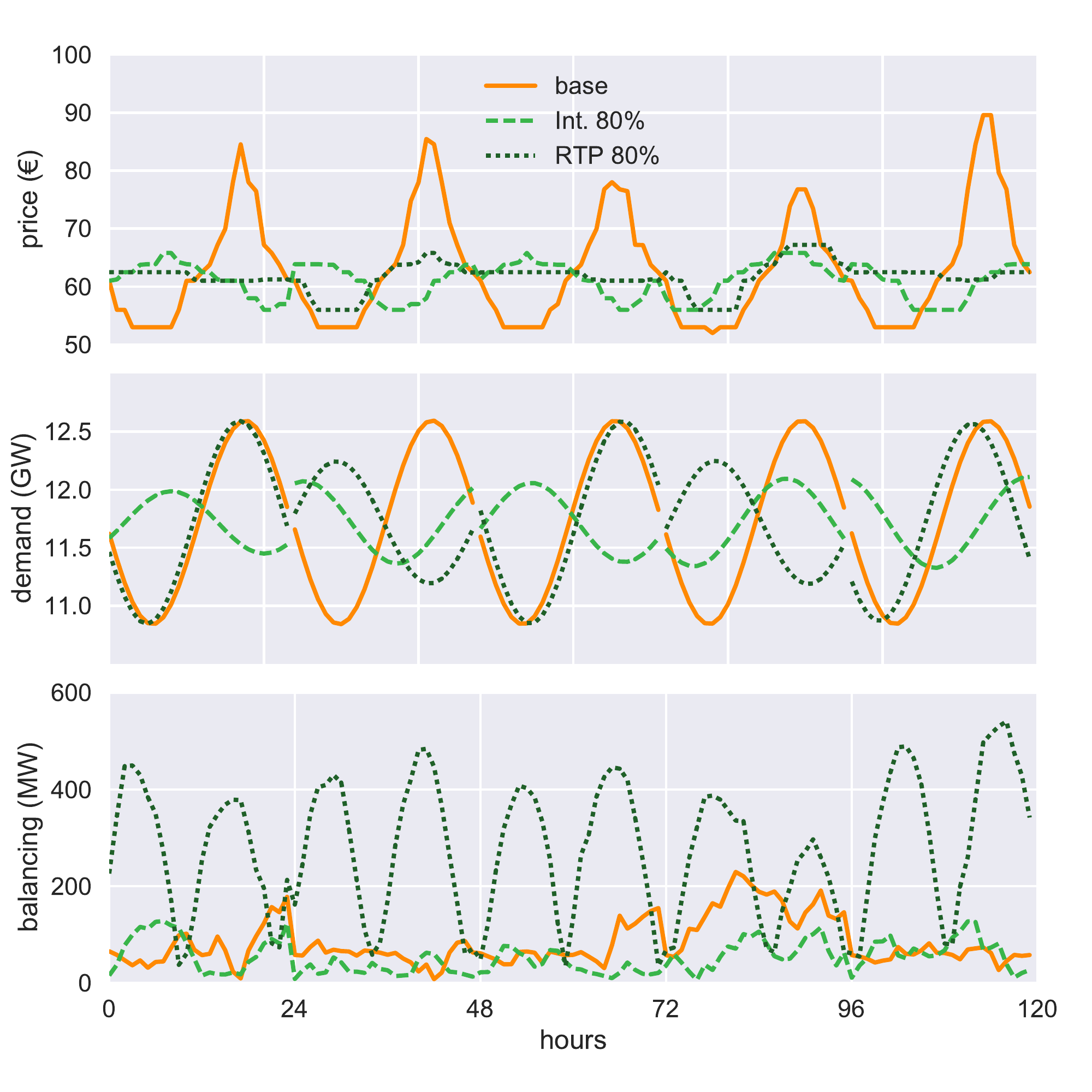}\\
 \caption{\label{INT80_vs_RTP80}A direct comparison between the the base case scenario with no flexibility and two with high amount of flexible load, one under each regime. The balancing plot shows up plus down balancing power needed. Both price and demand plots are not shown with a zero based y-axis. Note how in the RTP regime price and peak power consumption are now occurring at different times, and consumption is very erratic. The current implementation does not consider continuity beyond a single day, therefore jumps between two days can appear, without any impact. }
\end{figure*}

\begin{figure*}
\centering
  \includegraphics[width=\textwidth]{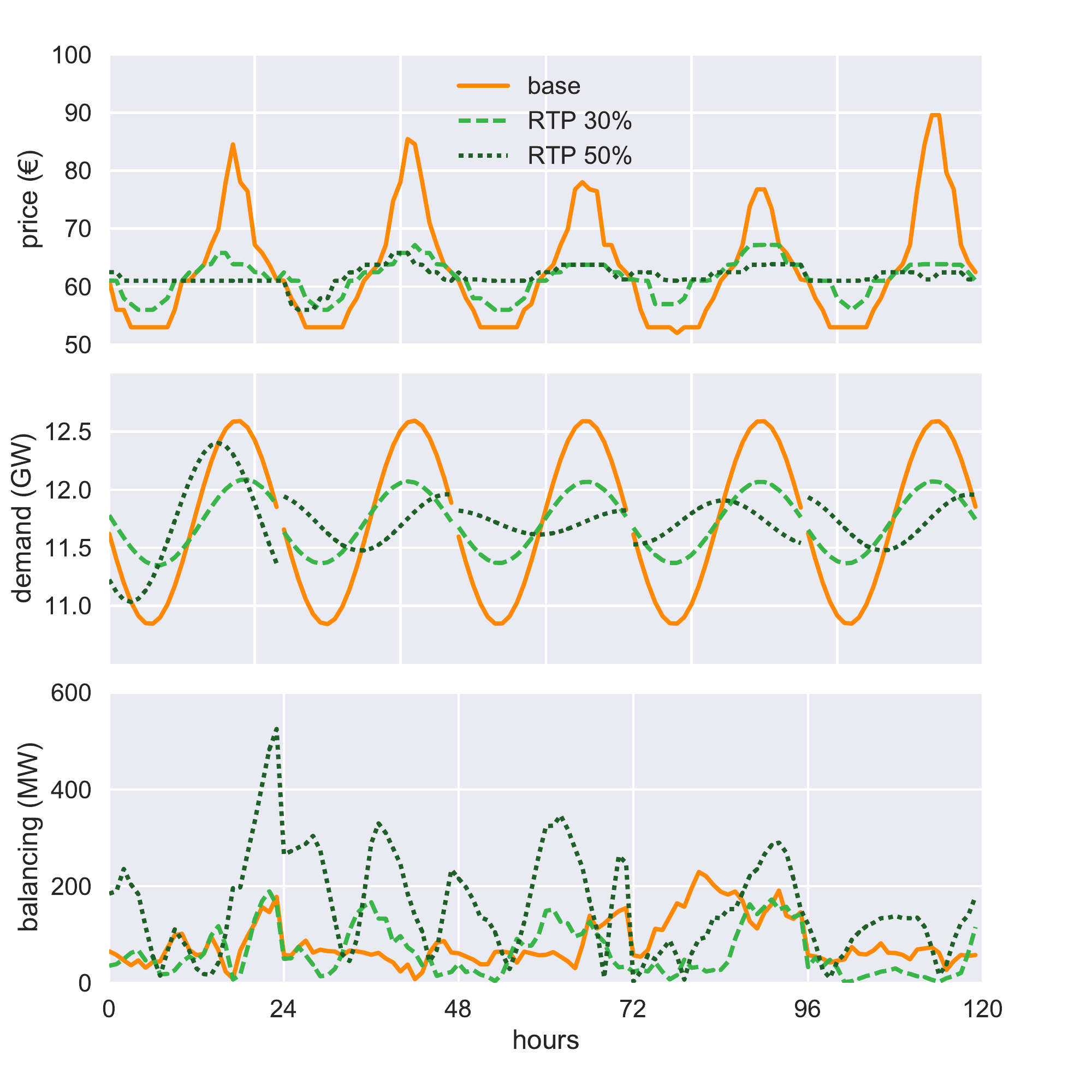}\\
 \caption{\label{RTP_30_50}A direct comparison between the the base case scenario with no flexibility and a two cases with 30\% and 50\% flexibility under the RTP regime.}
 
\end{figure*}

\begin{figure*}
\centering
  \includegraphics[width=\textwidth]{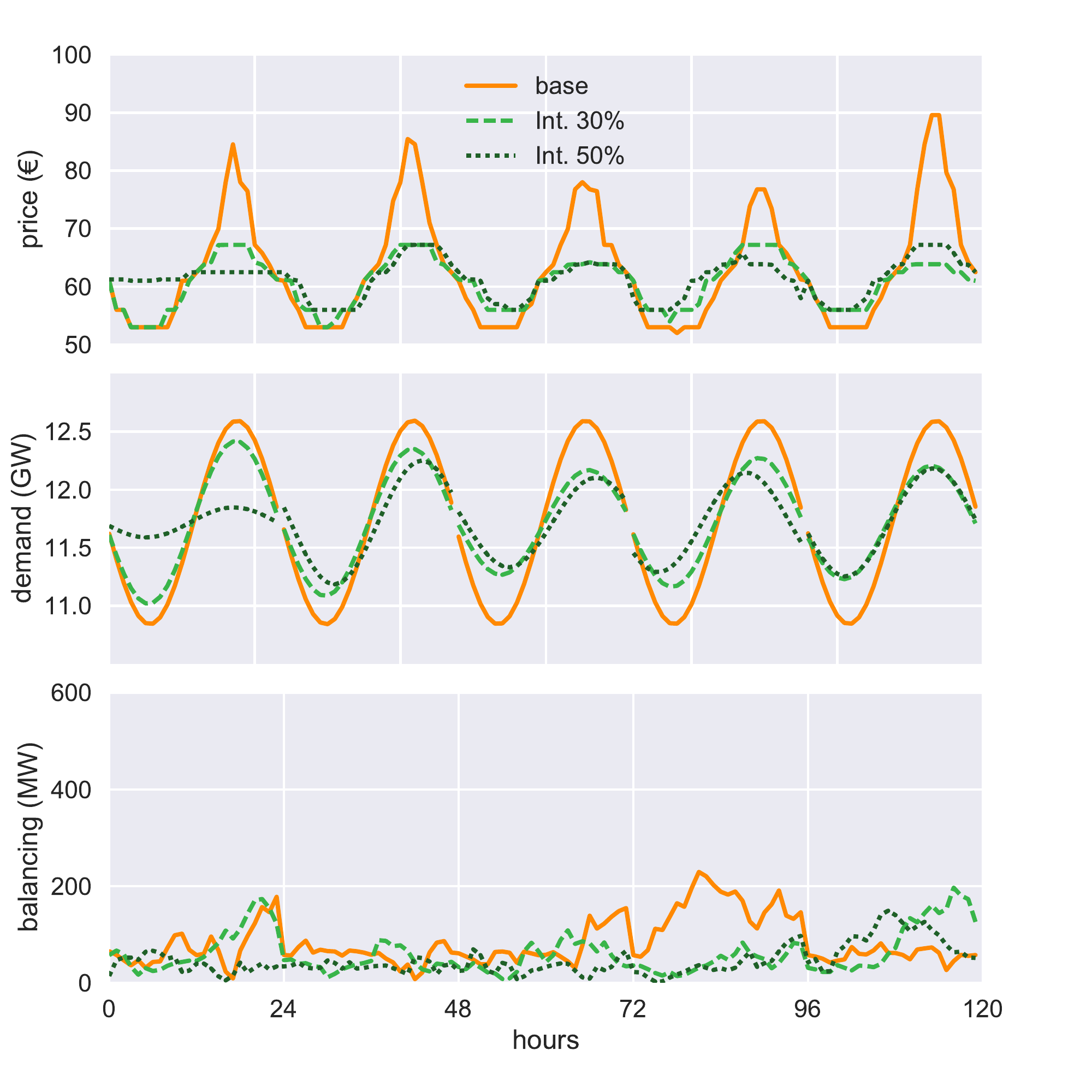}\\
 \caption{\label{int_30_50}A direct comparison between the the base case scenario with no flexibility and a two cases with 30\% and 50\% flexibility under the integrated regime.}
 
\end{figure*}

\begin{figure}
	\centering
	\includegraphics[width=0.6\columnwidth]{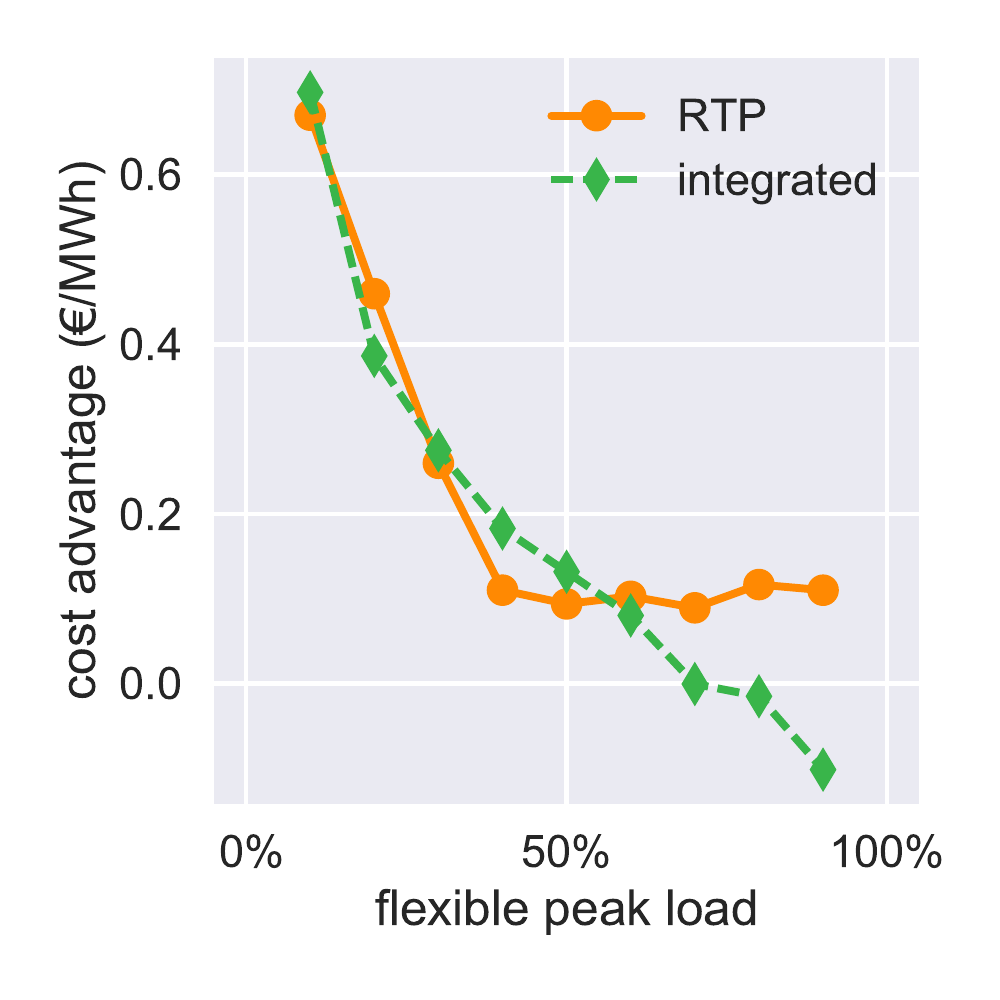}\\
	\caption{\label{group_comparison}A comparison between the two user groups (flexible users vs. non flexible users) in both regimes (RTP: solid orange, integrated via EXGs: dashed green). The solid line shows the relative cost advantage of flexible users in the RTP regime, while the doted line shows the advantage of flexible users in the integrated regime. Note: this is a relative comparison inside a given regime, overall costs are compared in Fig.\ref{savings_split}. Flexible users always pay less then non flexible users, despite the steep rise of system costs in the Real-time Price (RTP) regime.}
\end{figure}

The central part of our results can be seen in Figs. \ref{savings_split}-\ref{group_comparison}, showing that, above a certain ratio of flexible users in the system, flexibility guided by real-time price (spot price) dramatically increases the total electricity costs.
In other words, the savings that price-reactive users are expected to bring do not realized.
However, if we integrate the flexibility into the market with exclusive groups, the savings can be sustained regardless of the ratio of flexible users.
This effect results mainly from the consideration of the physical reality, captured by the balancing market. 

This can be seen in Fig. \ref{savings_split} specifically in the center plot where the usage related costs (i.e. the consumers' individual tariff based on the spot prices) remain low.
The system costs resulting from the balancing, however, quickly rise after a certain ratio of flexible users, as shown in the right part of the plot.
This behavior can be explained as follows.
The flexible users try to avoid high spot prices from the day ahead market, as they are billed directly according to them when they adopt the real-time pricing policy.
After a certain ratio is exceeded, the usage pattern from the flexible users influences when high prices occur.
This then leads to the peculiar situation that the users will never do what they were forecasted to do; the demand predictions by the utilities are never realized due to flexible-optimizing consumers.
If the utilities expect users have their peak consumption during a certain time of the day, their bids will increase prices during that hour.
This high price will lead to users avoiding these very hours.
This in turn means that down-balancing is needed in the high-price hours, while up-balancing is needed for other low priced hours.
It is exactly this balancing that will drive the system costs up.

A more detailed view into the simulation can be seen in Fig. \ref{INT80_vs_RTP80}, which shows price, demand, and balancing power for five days of simulation under integrated and RTP regime.
It illustrates how price and demand are misaligned in the RTP regime, leading to a high need of balancing power.
Another interesting observation can be made based on Fig. \ref{INT80_vs_RTP80}: even though it would be possible to create a flat load curve, eliminating all peaks, this is not the chosen solution by the market algorithm. 
This is due to non-linear effects in the merit order curve.
Above a certain level of consumption, power plants with higher costs need to run. 
It can therefore be cheaper to utilize them fully, once they are turned on but keep them off during the rest of the day.
This situation might change with available power plants.

Figs. \ref{RTP_30_50} and \ref{int_30_50} present similar results to Fig. \ref{INT80_vs_RTP80} considering RTP and integrated regime with 30\% and 50\% of flexible users; these plots corroborate our previous argument.
It is also worth stating that, while it is possible to argue that the changes caused by RTP may be predicted in the demand forecast, this does not change the previously described dynamics.
A difference in price is assumed to lead to a shift of demand.
Once this shift in demand is big enough to drive day-ahead price changes, users see every day a different ``real-time'' price curve based on their past behavior even including their expected changes.
As the actual consumption happens one day after the daily spot price is formed, the predictions that create that price cannot be realized as the spot (day ahead) price acts as the ultimate guide of the consumers' consumption pattern of that day whose price was formed one day before.
All in all, the system works with self-referential predictions, which imply that the predictions cannot be realized.

This unfortunate (idealist) deployment of flexibility via real-time (spot) prices, instead of creating a \red{self-fulfilling} prophecy, creates its converse: a never-fulfilling prophecy.
But, by knowing the systemic mechanism that creates this outcome, the justification to use EXGs as a tool to integrate  flexibility in large scale without creating extra costs becomes straightforward:
as far as the schedule of the flexible demand is set in the day ahead market, there will be no excessive need for balancing power so the system costs stay low.

Intuitively, one might expect that, from the perspective of the individual user, it is not beneficial to become flexible once the system under a real-time pricing regime reaches the point where costs start to rise again. 
In other words, one might expect that the system will automatically stay close to this optimal point.
As can be seen in Fig. \ref{group_comparison}, this is not necessarily the case.
The plot shows the cost advantage of flexible users (the ones who are able to shift their peak usage) over ordinary users considering: the Real-time Price (RTP), and the integrated regime where exclusive groups (EXGs) are used.
We can see that the cost advantage for flexible users in the RTP regime never goes below, or even close to, zero.
This in turn means that it is always beneficial to be a flexible user, since such users always pays less then the ordinary ones, regardless of the system configuration.
Assuming that there is no lower limit for switching, which is given by the return of investment for becoming a flexible user, this would drive the system to the worst possible state.
This fact resembles the well-known tragedy of the commons problem \cite{ostrom2008tragedy}.

While the same incentives still exist in the integrated regime, they lead to a different outcome: it accommodates the users' dynamic without increasing the total costs.
In any case, it is worth noting that a direct comparison between flexible and normal users under the two different regimes within the same system cannot be drawn from this graph.
Our analysis considers that, depending on the setting considered, flexible consumers are exclusively RTP or integrated, they do not coexist in the simulation; i.e. we  only consider scenarios: (i) RTP flexible users with normal users, and (ii) integrated-regime flexible users with normal users.
Therefore Fig. \ref{group_comparison} do not present any comparison between the two regimes (this conclusions can be drawn form Fig. \ref{savings_split} already).
Instead, Fig. \ref{group_comparison} illustrates the internal relation between the two different kind\red{s} of users in a specific regime.
While a large amount of flexibility increases overall costs in the RTP regime, it remains  beneficial for individual consumers to become flexible in comparison to the normal users.
In the integrated regime, we see the gains are relatively even for high ratio of flexible users with diminishing individuals gains.

\begin{figure*}
\centering
  \includegraphics[width=\textwidth]{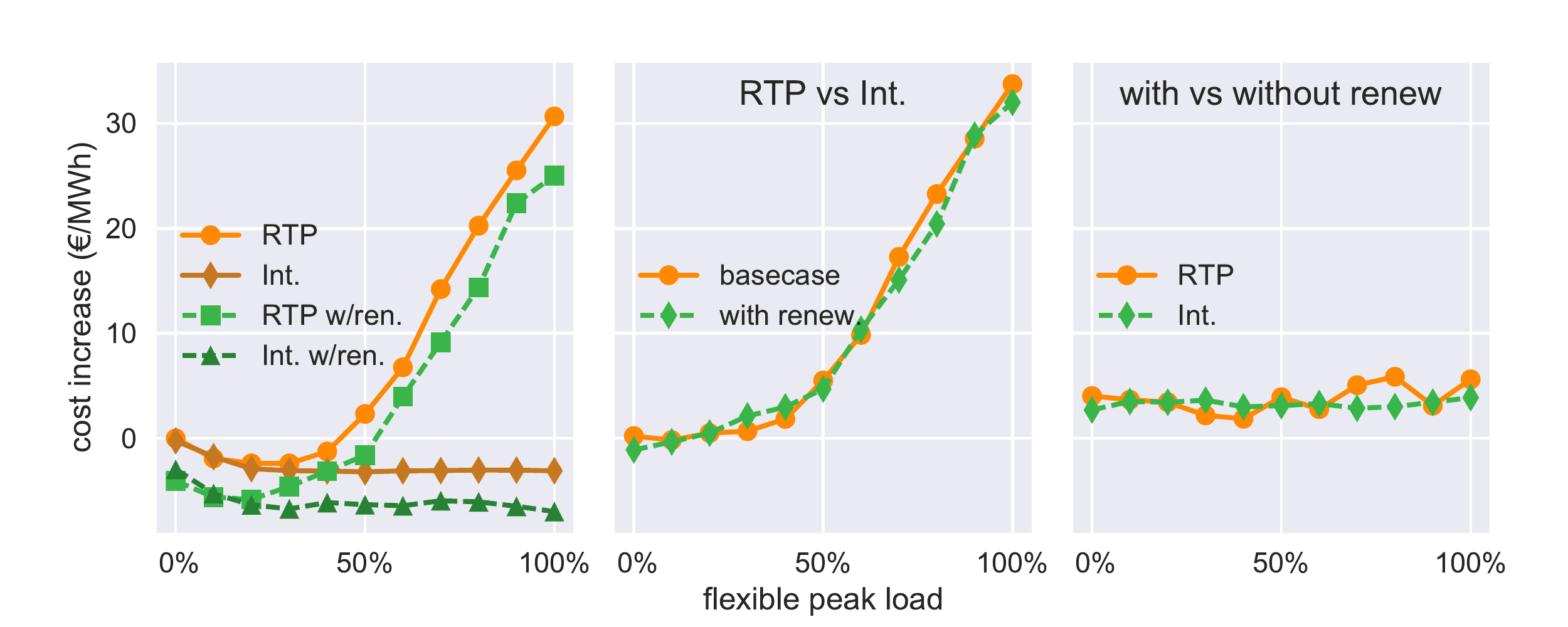}\\
 \caption{\label{renew}Effects of an intermittent renewable power available randomly during each simulation day (up to 2 GW peak) with a certain forecast error. Left: the cost increase vs. the ratio of flexible users for RTP and integrated regimes with and without intermittent renewable sources. Center: the cost increase of the RTP regime over the integrated without and with renewables. Right: the cost increase of the integrated regime with and without renewables for the same regime. The plot in the center is computed as the the gap between the curves with and without renewables considering from the left plot, while  the right plot is the gap between the base cases and the ones with renewables.} 
\end{figure*}

In Fig. \ref{renew}, we simulated the effects of adding an intermittent renewable source to the system.
More specifically, we added a producer that would provide up to 2 GW at random periods during the day.
The duration and shape of the periods are also random.
Additionally, the realized output differs from the forecasted one.

We identify the following effects of the presence of such a source. 
During the times when renewable power is available, the day ahead market prices fall, as less conventional power needs to be dispatched.
Then, more balancing power is required to cover the differences between renewable forecast and realized output. 
For the system under real time pricing, we expect a bigger impact, i.e. a drastic cost increase with a lower ratio of flexible users, as the utilities do not predict based on the availability of those renewable sources.

The results in Fig. \ref{renew} corroborate this view, as we see a bigger cost increase for a share of 30\% of flexible users. 
When comparing the systems with and without renewable energy in the respective regimes, we verify the cost advantage that the renewable sources bring is sustained in the integrated regime, while it varies noticeably in the RTP one.
Overall, the cost reduction with the addition of renewables is greater than the gains through flexibility.
Additionally, the flexibility does not provide bigger benefits in the renewable system.

\section{Discussion}
\label{sec-disc}

Even though variable pricing schemes are implemented for quite some time, only very recently researchers started to identify possible problems that arise when they become massively used as a guide for shifting flexible loads. 
For example, problems related to market instabilities were presented in \cite{Roozbehani:2012gn}, while instabilities of the physical grid due to these effects were discussed in \cite{Belonogova:2013cp}.
In \cite{Krause:2015hp} undesirable demand synchronization between independent agents have been demonstrated with a simple agent-based model and external price signals.
With the specific example of electric vehicles, the work of \cite{Dallinger:2012ib} shows a type of ``avalanche effect'' where the charging times of all vehicles synchronize in low price times.

With this work, we expect to illustrate problems in the same realm, but with a different -- and quite specific -- working mechanism as well as an already existing solution to this very problem.
We consider the electricity system as a whole, so that each element influences each other, following the regulatory structure defined by the market as it currently exists and the customer pricing schemes already in place in many EU countries.

As any other model, ours is also built upon a number of simplifications\cite{Sterman:2002ef}; we in fact consider the proposed model as the simplest yet functional version to demonstrate the effects previously discussed.
Several of these simplifications, however, should be examined to better understand the validity of the results.
Probably, the clearest simplification is the assumption of a sinusoidal load curve that is only flexible in one dimension.
In reality, the load curve is composed out of a number of devices installed in a household or commercial building.
Each device has a very distinguished consumption pattern and flexibility.
Therefore the optimization inside a household would change the load curve in multiple dimensions and possibly converge to a different optimum for every new day.
We do not expect, though, any fundamental changes to the results if more details and degrees of freedom would be considered in the load modeling.
Even in the extreme case where each consumer would be able to freely shape his consumption (for example, with sufficient local storage), it is unlikely that the system would converge to a more favorable state in the RTP regime. 
Small differences in hourly prices would always create incentives to load curves that are not flat as users try to gain by utilizing low price hours and avoiding higher priced ones.

On the other hand, it might be plausible to assume that some of the effects will be enhanced by a more detailed load model.
For instance, if a single flexible device like a household's dishwasher is optimizing its own usage following the day-ahead prices, it then tries to minimize costs by allocating its activity accordingly.
Assuming that this phase is reasonably similar for all dishwashers, the optimization process would lead to all dish washers to be activated during the same hour of the day.
In fact, it is very likely that all dishwashers would turn on at same second of the same hour to optimally use the time where energy is cheap according to the real-time tariff.
This would lead to an extremely steep jump in consumption and a concentration of the need for balancing energy, possibly (and most probably) increasing system costs even more.
Similar effect has been described for heating in \cite{Belonogova:2013cp}.

The last example also illustrates another assumption concerning the load modeling.
We consider that there are no costs involved in shifting the energy usage, which is contrary to what is often presumed.
In reality, this might differ from appliance to appliance, where energy storage will have opportunity cost or cycle costs, while a dish washer can just turn on whenever it wants within a certain predetermined time frame. 
However, we believe most flexible devices will be more like a dishwasher and not like a battery, as it seems unlike that consumers would be willing to trade their comfort for money, given how little money can be saved \cite{Gottwalt:2011do,Kim:2011fy}.
Even for the charging of an electric car battery, the user might only be concerned about the battery being fully charged in the morning and afternoon hours for the trip from and back to work.
Hence, the optimization would deal less with opportunity costs than the actual need.
Likewise, consumers might be willing to have the set point of their heating changed up or down by one or two degrees as this is mostly unnoticeable, but no changes in how warm the house feels would be tolerated as the savings will, at most, sum up to a single digit amount of € per day.

Assuming no shifting costs may have important implications.
First, it might lead to a more volatile system behavior as it might be profitable to exploit even very small differences in price.
Second, it might result in the lack of a damping factor for the system to converge to a stable price equilibrium (if a more analytical approach is used).
For our results, both of these factors should not be of concern, as the price differences are rather big and the shift of user behavior only occurs once a day, so damping of oscillating user behavior during the day is not an issue here.

Furthermore, we believe almost all changes in consumption will happen through automation, not from  changes in human behavior or routines (i.e. the peak hours are closely related to the working day so the human-driven consumption patterns follow the daily routine: waking up, eating, showering and so on).
This implies that, if changes in consumption are driven by human behavior, they will be slower and less reactive to short-term changes. 
If an appliance is automated to react to a certain external signals, it might not make any difference for the user what that signal is. 
All in all, if the dishwasher turns on driven by pricing information or a remote control signal, it is indistinguishable from the user perspective, as he or she would not have direct influence on either of them. 

\red{Another grid element that exhibits many of the above properties is the electric vehicle during charging. Since charging is flexible in its starting time and also in duration (and therefore peak power consumption), it can have massive impacts especially on distribution grids\cite{Sharma:2014jb,Xu:2014hl,Auer:2017wn}. As pointed out in \cite{GarciaVillalobos:2014hr} a centralized solution to coordinate charging that assures all network parameters are accounted for is most likely needed, which would fits with the EXG integration approach.}

An additional concern might be related to the ability of utilities to forecast user behavior.
At  first, one might assume that  it is possible to better forecast how users react to prices with a more sophisticated learning algorithm.
While this might be true to a certain extent,  several assumptions of our model prevent any big gains from this perspective.
The important part to consider is the feedback loop which happens between markets, users and utilities.
This feedback loop is not closed within any given day as utilities have to make their predictions about user behavior before the market closing time, which is before the beginning of the operational day.
Users, however, can alter their behavior based on the outcome of the market phase during the operational day.
In other words, users can always change their behavior after the utilities made public what they forecasted.

Another way for the utility to improve the outcome appears to be price-dependent orders.
The idea is the following: a utility can learn the flexibility of its users and then bid a price-dependent order into the market, which reflects that, for high prices, less energy will be consumed and vice versa.
We did not include these in the modeling as they offer no significant improvement to resolve the situation explained before.
The hourly price-dependent order does not correctly reflect the shift in demand.
If the price is low, more energy will be bought, but this does not necessarily correspond to a reduction of demand during another part of the day.
This mechanism works well for industrial consumers that might increase production of a good if prices are low or rely on internal power generation if prices are high enough, as these processes are typically without temporal interlocking (e.g. production can be increased during one hour without the need to decrease during another hour).
Furthermore, the price-dependent order does not work with the assumptions made here, where a shift in demand might not be based on the absolute but rather the relative price (like the automated dishwasher that just looks for the cheapest hour, not for a specific price).
\red{There are numerous processes like this in the industry, the private sector and in households, which will only run once, and cannot run again even if cheaper power would be available.}
It is important to point out that the model presented here also demonstrates that flexible users in the RTP regime can bring benefits but only when their scale is limited, as can be seen in Fig. \ref{savings_split} and in Fig. \ref{RTP_30_50}.
In this case, the price-dependent behavior of flexible users can be correctly forecasted by the utilities and all the assumed benefits of RTP combined with flexibility shall be realized.

It appears that this problem might be solved by using a price that is closer to real-time than the day-ahead price, namely the intraday price or the balancing price.
While the instabilities of such system have been demonstrated in, for example, \cite{Roozbehani:2012gn}, there are also logical arguments against it.
Every market stage is based around binding settlements.
If a bid or an offer was accepted, it is a binding settlement to either produce or consume a certain amount of power.
In the real-time pricing regime, there is no binding agreement.
This is especially problematic since the price is the only information available; yet there is only a finite amount of power that can be produced for that price -- although, from the customer side, it appears as if the amount is unlimited.

This is of course not true in any market.
It is not possible to buy all the stock of a company for its ``stock price".
The stock price is either shown as the gap between the bids and offers or the last successful trade.
Neither of those are the price at which you can currently buy shares.
Similarly, one cannot buy unlimited amounts of butter for the price listed in one specific supermarket: one can only buy the quantity in its storage.

It should be also noted that the remote control signal to a device can take many forms. 
For example, the utility may have run an internal auction with all participating appliances or may have an internal pricing scheme.
Similar ideas have been portrayed in \cite{Garcia:2017hl,YOUSEFI20115716}.
Either way, the utility needs to somehow schedule the automated devices, at least up to a certain extent.
Otherwise, the issues regarding synchronization and its undesirable consequences may emerge.

Given these simplifications and limitations, we are aware that our current results cannot be used as a predictor for how big a RTP system can be safely scaled up. 
However, it shows the systemic drawbacks of this widely promoted solution and the advantages of a market integrated solution (if implemented properly).
We further believe that some critical misconceptions about the currently proposed RTP solution should be considered in the discussion about this topic, as follows:
\begin{itemize}
\item Contrary to what the name suggests, prices are not updated in real time as we commonly understand it, but rather are determined up to 36 hours before the operational hour.
\item Prices reflect binding settlements from the market place. A change of the agreed schedule breaks the contract.
\item Since order books are closed, it is unknowable how much power is available for a given price, making optimal scheduling impossible.
\end{itemize}

\red{Besides, it is worth pointing out that studies of demand response for household customers and RTP have typically found  little benefit for the customer\cite{Connell:2014fg,Kim:2011fy}. In a relevant case-study of the implementation of RTP in Chicago \cite{Allcott:2011gs}, the authors found that the cost of the metering infrastructure is likely to exceed the benefits gained efficiency: ``More importantly, although there are significant uncertainties, these efficiency gains do not appear to overwhelm reasonable estimates of the cost of conserving energy or installing advanced metering infrastructure. These results do not make a strong case for optional or population-wide residential real time pricing."}

As final point of the discussion we would like to point out a drawback of the market integrated solution based on EXGs.
Since the matching of bids and orders is essentially an optimization problem, e.g. selecting a set with the highest social welfare considering several border conditions, like transmission constraints, is a big computational task.
The computational effort needed to solve this problem is already quite big for the European market:
all orders (Nord Pool and EPEX combined) are put in only one order book so that transmission constraints can be optimized for.
The size of the optimization problem leads to a lack of transparency of the solution found by the market, which is becoming a growing concern \cite{PCRStatusUpdate:2016tl}.
Adding more of the so-called complex orders (like the proposed exclusive block orders) would dramatically increase the computational complexity of the optimization problem, consequently decreasing the transparency of the market solutions.

\section{Conclusion}
\label{sec-concl}

In this paper, we presented an extension of the agent-based spot and balancing market model proposed in \cite{kuhnlenz2017agent} to include flexible consumers. 
The model was calibrated as a representation of an electricity system similar to the Finnish one, as part of Nord Pool.
From this model, we analyzed the effects of two distinct regimes of demand-response, namely ``real-time pricing'' and integrated.
In the first one, consumers optimize their consumption based on the tariff scheme that charges them according to the spot market prices for the day.
In the second one, the consumers' flexibility is integrated into the market by aggregating it on the utility level and then scheduling the flexible usage following the profile accepted by the market.

We showed that above a certain ratio of flexible consumers inside the market area, the real-time pricing regime leads to a situation where the utility is unable to correctly predict the realized consumption of its users.
We argue that this inability is a systemic feature, which can be corrected the integrated solution using existing market products and a more direct control of the flexibility.
Our results also indicate that the presence of intermittent renewable sources can decrease the electricity prices, but it does not change the qualitative behavior of the two studied regimes.
The proposed model is open-source and available on-line at \cite{sourceCode}.

\section*{Acknowledgements}

\noindent
This work is partly funded by Strategic Research Council/Aka BCDC Energia (n.292854) and by the European Commission through the P2P-SmarTest project (n.646469).

\clearpage


%

\begin{figure}
	\centering
	\includegraphics[width=0.6\columnwidth]{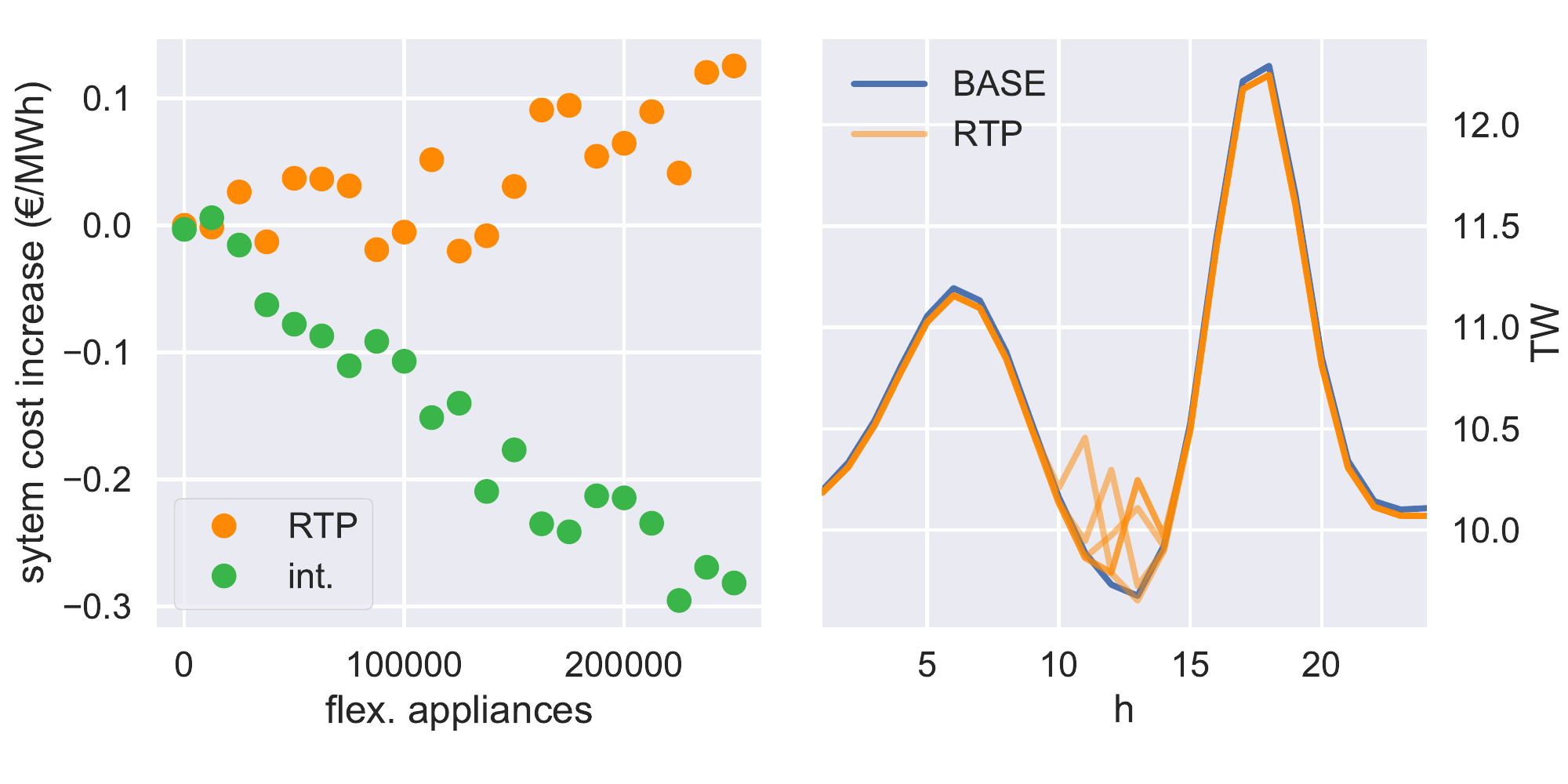}\\
	\caption{\label{app-figure}Comparison of small flexible loads (e.g. smart dishwashers) with a power consumption of about 3kW for about 60 min. On the left we see the results of coordinating these loads via EXG and market integration or RTP. On the right we see the base-load curve and the "jumping" of loads in the RTP regime, where all appliances try to run during the cheapest hour.}
\end{figure}

\appendix
{
\section{Modeling of Appliances}\label{app-model}

To evaluate the consistence of the proposed simplified load profile, we design an experiment to test whether the downsides of RTP still valid when more realistic load curves are used.
We set a system in a quite different manner compared to the setup employed in the main body of this paper.
Instead of sinusoidal load curves we rely on the following model for the demand: Every utility has one agent that follows a load curve with two peak during the course of a day, one during morning and one during evening hours, as can be seen in Fig. \ref{app-figure} on the right. The sole task of this agent is to represent some form of base-load.
Each utility has a set of agents that represent appliances, which can either be normal or optimizing, and each utility has the same amount of optimizing and normal appliances. Every appliance has to run once during a day for about one hour and consumes a given amount of power during that time \cite{Srikantha:2012gi,Saele:2011hi}. The duration and amount of power differ slightly for each appliance. For the normal appliances the time when they are on is distributed during the day with a distribution mirroring the demand curve, so more appliances are running during the morning and evening hours.

As before, the optimizing appliances can either be under RTP or integrated into the market via profile orders handled by the utilities. During the simulation, we altered the amount of power that is consumed by either normal or optimizing appliances proportionally, so that as a whole they would represent about 250.000 appliances with a power demand of 3kW each. 

As can be seen in Fig. \ref{app-figure} on the left, the results are comparable to our main findings, and can be understood via the same concepts. As seen on the right side of Fig. \ref{app-figure}, the appliances under RTP always try to jump to the cheapest hour during mid-day. The utilities try to forecast to which hour they will move. However this very prediction is bid into the market and will therefore alter what hour will be the cheapest one. Consequently, the appliances under RTP are constantly producing a need for additional procurement that comes at higher costs. This ultimately drives up the system costs.

In the integrated regime, the market can coordinate when the appliances will run and procure the needed power beforehand, so no additional costs for power procurement after the day-ahead market will occur. Of course, the effects in this setup are much smaller than the effects shown in the paper since the influence of the appliances on the peak load times are smaller. However, this shows that just a fleet of 250.000 smart dishwashers might be able to create negative results for the whole system if not controlled correctly.

We would like to point out that this research path needs to be further developed to have a better understanding of the effects of different load curves, of different (group of) appliances and of the inclusion of more boundary conditions\cite{Zhong:2015ku,Zhong:2015hj} on the flexibility and its management. However, this is outside the scope of this article, which focuses on understanding the underlying effects and the problem in general. 
}

\section{Model Description}\label{model-description}
In the following, we present the pseudo-code representation of the proposed model. Some details of the implementation necessarily need to be omitted for simplicity; these details are available in the source code itself, available on-line at \cite{sourceCode}. For example, we choose to exclude the details of the calculation of the balancing costs as it is a rather long process. This, however, just follows the accounting rules of the Nord Pool system.
We advise to start with Algorithm \ref{main} as it is the main loop iterating over the days of the simulation.

\begin{algorithm}
\caption{Run For Days}\label{main}
	\begin{algorithmic}[1]
		\For{$day \gets 1, days$ }
        	\State system.startDay() \Comment{Internal setup}
            \State $prices,schedules,profiles \gets $ system.runMarket()
            \State system.setSchedule($schedules$)
            \State system.setProfiles($profiles$)
            \State system.updatePostMarket()
            \State $balancingPrices \gets $ system.performBalancing()
            \State system.nextDay() \Comment{Used for state reset of agents and cost calculation}
        \EndFor
	\end{algorithmic}
\end{algorithm}

\begin{algorithm}
\caption{Run Market}\label{market}
	\begin{algorithmic}[1]
    	\Procedure{runMarket}{}
    	\State $supply$ = system.getSupply() \Comment{Collect offers from producers}   
        \State $profiles \gets []$
        \For{$util$ in $utilities$} \Comment{Collect profiles from utilities}
        	\State $profiles$.add($util$.profilesForDay())
        \EndFor
        
        \State $selection,best\_selection \gets $ system.lastSelection
        \State $value,best \gets \textrm{eval}(selection,supply,profiles)$
        
        \For{$i \gets 0, n$}
        	\State $T \gets \textrm{ANNEAL\_TEMP} \slash (i+1)$ 
            \State $move\_vec \gets$ randomVector()
            \State $new\_selection \gets selection + move\_vec$
            \State $new\_value \gets \textrm{eval}(new\_selection,supply,profiles)$ 
            \If{$ \exp(new\_value - value) \slash T \textrm{ OR } new\_value > value  $}
            	\State $selection \gets new\_selection$
                \State $value \gets new\_value$
            	\If{$new\_value > best$}
                	\State $best\_selection \gets new\_selection$
                 	\State $best \gets new\_value$
               	\EndIf
            \EndIf
        \EndFor
        \State $prices,schedule \gets \textrm{createSchedule}(profiles(best\_selection),supply)$
	\State \Return prices,schedule,best\_selection
    \EndProcedure
    \end{algorithmic}
\end{algorithm}

\begin{algorithm}
\caption{Evaluate Profile}\label{eval}
	\begin{algorithmic}[1]
		\Procedure{eval}{selection,supply,profiles}
        	\State $profile \gets profiles(selection)$
            \State $surplus \gets 0$
            \For{$hour \gets 0, 23$}
            	\State $price,schedule \gets \textrm{priceForDemand}(profile(hour),supply(hour))$
                \For{$item \gets schedule$}
                	\State $surplus \gets surplus + item.accepted \cdot{} (price - item.price)$
                \EndFor
                \State $surplus \gets surplus + supply(hour) \cdot{} (3000 - price)$
            \EndFor
        \EndProcedure
    \end{algorithmic}
\end{algorithm}

\begin{algorithm}
\caption{Price For Demand}\label{priceForDemand}
	\begin{algorithmic}[1]
		\Procedure{priceForDemand}{demand,supply}
			\State $supply \gets \textrm{sort}(supply)$
            \State $available \gets 0$
            \State $schedule \gets []$
			\For{$offer \textrm{ in } supply$}
            	\State $available \gets available + offer.power$
                \State $price \gets offer.price$
                \State $schedule.add(\textrm{ScheduleItem}(offer))$
                \If{$available \geq usage$}
                	\State \Return $price,schedule$
                \EndIf
            \EndFor
        \EndProcedure
    \end{algorithmic}
\end{algorithm}

\begin{algorithm}
\caption{Create Schedule}\label{createSchedule}
	\begin{algorithmic}[1]
		\Procedure{createSchedule}{profile,supply}
            \State $schedule \gets []$
            \State $prices \gets []$
			\For{$hour \gets 0,23$}
            	\State $schedH,priceH \gets \textrm(priceForDemand)(profile(hour),supply(hour))$
            	\State $schedule.add(schedH)$
                \State $prices.add(priceH)$
            \EndFor
        \State \Return $prices, schedule$
        \EndProcedure
    \end{algorithmic}
\end{algorithm}

\begin{algorithm}
\caption{Update Post Market}\label{postMarket}
	\begin{algorithmic}[1]
		\Procedure{updatePostMarket}{}
			\For{$user \textrm{ in } allUsers$}
            	\State $user.updatePostMarket()$
            \EndFor
        \EndProcedure
    \end{algorithmic}
\end{algorithm}

\begin{algorithm}
\caption{User Update Post Market}\label{postMarketUser}
	\begin{algorithmic}[1]
		\Procedure{updatePostMarket}{}
			\If{isOptimizer AND isRTP}
            	\State optimizeUsage()
            \ElsIf{isOptimizer AND hasProfiles}
            	\State updateUsage()
            \EndIf
        \EndProcedure
    \end{algorithmic}
\end{algorithm}

\begin{algorithm}
\caption{Balancing Part 1}\label{balancing1}
	\begin{algorithmic}[1]
		\Procedure{performBalancing}{}
			\State $demand \gets $ system.minuteDemand() \Comment{actual demand of users with minute resolution}
            \State $production \gets $ system.minuteProduction() \Comment{actual production of producers with minute resolution}
            \State $diff \gets demand - production$
            \State $diff15 \gets \textrm{sum15}(diff)$ \Comment{Sum differences for 15 minute slots}
            \State $prices \gets \textrm{list}(96)$
            \For{$t \gets 0,95$}
            	\If{$diff15(t) > LIMIT$}
                	\State $offers \gets \textrm{getUpOffers()}$)
                    \State $demand \gets diff15(t)$
                    \For{$offer \textrm{ in } offers$}
                    	\State $demand \gets demand - offer.power$
                        \State $prices[t] \gets offer.price$
                        \If{$demand > 0$}
                        	\State updateSchedule($offer,offer.power$)
                        \Else
                        	\State updateSchedule($offer,offer.power + demand$)
                            \State break
                        \EndIf
                    \EndFor
                \EndIf
                \algstore{bkbreak}
                
    \end{algorithmic}
\end{algorithm}
                
\begin{algorithm}
\caption{Balancing Part 2}\label{balancing2}
	\begin{algorithmic}[1] 
    \algrestore{bkbreak}
                \If{$diff15(t) < -LIMIT$}
                	\State $demand \gets diff15(t)$
                	\State $offers \gets \textrm{getDownOffers()}$)
                    \For{$offer \textrm{ in } offers$}
                    	\State $demand \gets demand + offer.power$
                        \State $prices[t] \gets offer.price$
                        \If{$demand < 0$}
                        	\State updateSchedule($offer,offer.power$)
                        \Else
                        	\State updateSchedule($offer,offer.power - demand$)
                            \State break
                        \EndIf
                    \EndFor
                \EndIf
            \EndFor
            \State $diff60 \gets \textrm{sum60}(diff)$ \Comment{Sum differences for 60 minute slots}
            \State $priceH \gets list(24)$
            \For{$t \gets 0,23$}
            	\If{$diff60(t) > LIMIT$}
                	\State $priceH[t] \gets \max(price[t\cdot4:t\cdot4+4])$
                \ElsIf{$diff60(t) < -LIMIT$}
                	\State $priceH[t] \gets \min(price[t\cdot4:t\cdot4+4])$
                \EndIf
            \EndFor
            \State \Return $priceH$
        \EndProcedure
    \end{algorithmic}
\end{algorithm}

\FloatBarrier

\section{}
\begin{longtable}{l|l|l|l|l|l}
\caption{System Configuration}\\
Idx. &  Capacity & Marginal Cost €/MWh & Reg. Factor & Reg. Update Factor & Min Run Factor \\\hline
1     & 232                  & 5.00                & 5\%               & 2                        & 0.1            \\
2     & 160.5                & 5.00                & 5\%               & 2                        & 0.1            \\
3     & 134                  & 5.00                & 5\%               & 2                        & 0.1            \\
4     & 128                  & 5.00                & 5\%               & 2                        & 0.1            \\
5     & 116                  & 5.00                & 5\%               & 2                        & 0.1            \\
6     & 105                  & 5.00                & 5\%               & 2                        & 0.1            \\
7     & 105                  & 5.00                & 5\%               & 2                        & 0.1            \\
8     & 90                   & 5.00                & 5\%               & 2                        & 0.1            \\
9     & 88                   & 5.00                & 5\%               & 2                        & 0.1            \\
10    & 85                   & 5.00                & 5\%               & 2                        & 0.1            \\
11    & 82.8                 & 5.00                & 5\%               & 2                        & 0.1            \\
12    & 76                   & 5.00                & 5\%               & 2                        & 0.1            \\
13    & 76                   & 5.00                & 5\%               & 2                        & 0.1            \\
14    & 76                   & 5.00                & 5\%               & 2                        & 0.1            \\
15    & 75                   & 5.00                & 5\%               & 2                        & 0.1            \\
16    & 75                   & 5.00                & 5\%               & 2                        & 0.1            \\
17    & 72.6                 & 5.00                & 5\%               & 2                        & 0.1            \\
18    & 72                   & 5.00                & 5\%               & 2                        & 0.1            \\
19    & 72                   & 5.00                & 5\%               & 2                        & 0.1            \\
20    & 67.3                 & 5.00                & 5\%               & 2                        & 0.1            \\
21    & 64                   & 5.00                & 5\%               & 2                        & 0.1            \\
22    & 48                   & 5.00                & 5\%               & 2                        & 0.1            \\
23    & 45                   & 5.00                & 5\%               & 2                        & 0.1            \\
24    & 43                   & 5.00                & 5\%               & 2                        & 0.1            \\
25    & 41                   & 5.00                & 5\%               & 2                        & 0.1            \\
26    & 40                   & 5.00                & 5\%               & 2                        & 0.1            \\
27    & 40                   & 5.00                & 5\%               & 2                        & 0.1            \\
28    & 39.3                 & 5.00                & 5\%               & 2                        & 0.1            \\
29    & 35.7                 & 5.00                & 5\%               & 2                        & 0.1            \\
30    & 35                   & 5.00                & 5\%               & 2                        & 0.1            \\
31    & 34.5                 & 5.00                & 5\%               & 2                        & 0.1            \\
32    & 32.8                 & 5.00                & 5\%               & 2                        & 0.1            \\
33    & 30                   & 5.00                & 5\%               & 2                        & 0.1            \\
34    & 28                   & 5.00                & 5\%               & 2                        & 0.1            \\
35    & 26.3                 & 5.00                & 5\%               & 2                        & 0.1            \\
36    & 25.1                 & 5.00                & 5\%               & 2                        & 0.1            \\
37    & 25                   & 5.00                & 5\%               & 2                        & 0.1            \\
38    & 20.5                 & 5.00                & 5\%               & 2                        & 0.1            \\
39    & 19                   & 5.00                & 5\%               & 2                        & 0.1            \\
40    & 19                   & 5.00                & 5\%               & 2                        & 0.1            \\
41    & 17.5                 & 5.00                & 5\%               & 2                        & 0.1            \\
42    & 16.5                 & 5.00                & 5\%               & 2                        & 0.1            \\
43    & 16                   & 5.00                & 5\%               & 2                        & 0.1            \\
44    & 14.2                 & 5.00                & 5\%               & 2                        & 0.1            \\
45    & 13.5                 & 5.00                & 5\%               & 2                        & 0.1            \\
46    & 12                   & 5.00                & 5\%               & 2                        & 0.1            \\
47    & 11.4                 & 5.00                & 5\%               & 2                        & 0.1            \\
48    & 8.8                  & 5.00                & 5\%               & 2                        & 0.1            \\
49    & 8.5                  & 5.00                & 5\%               & 2                        & 0.1            \\
50    & 7.8                  & 5.00                & 5\%               & 2                        & 0.1            \\
51    & 7.5                  & 5.00                & 5\%               & 2                        & 0.1            \\
52    & 7                    & 5.00                & 5\%               & 2                        & 0.1            \\
53    & 6                    & 5.00                & 5\%               & 2                        & 0.1            \\
54    & 5                    & 5.00                & 5\%               & 2                        & 0.1            \\
55    & 4.2                  & 5.00                & 5\%               & 2                        & 0.1            \\
56    & 4                    & 5.00                & 5\%               & 2                        & 0.1            \\
57    & 3.9                  & 5.00                & 5\%               & 2                        & 0.1            \\
58    & 3.8                  & 5.00                & 5\%               & 2                        & 0.1            \\
59    & 3.5                  & 5.00                & 5\%               & 2                        & 0.1            \\
60    & 3                    & 5.00                & 5\%               & 2                        & 0.1            \\
61    & 2.9                  & 5.00                & 5\%               & 2                        & 0.1            \\
62    & 2.8                  & 5.00                & 5\%               & 2                        & 0.1            \\
63    & 2.7                  & 5.00                & 5\%               & 2                        & 0.1            \\
64    & 2.7                  & 5.00                & 5\%               & 2                        & 0.1            \\
65    & 2.5                  & 5.00                & 5\%               & 2                        & 0.1            \\
66    & 2                    & 5.00                & 5\%               & 2                        & 0.1            \\
67    & 1.6                  & 5.00                & 5\%               & 2                        & 0.1            \\
68    & 1.5                  & 5.00                & 5\%               & 2                        & 0.1            \\
69    & 1.4                  & 5.00                & 5\%               & 2                        & 0.1            \\
70    & 1.4                  & 5.00                & 5\%               & 2                        & 0.1            \\
71    & 1.3                  & 5.00                & 5\%               & 2                        & 0.1            \\
72    & 1                    & 5.00                & 5\%               & 2                        & 0.1            \\
73    & 0.9                  & 5.00                & 5\%               & 2                        & 0.1            \\
74    & 0.8                  & 5.00                & 5\%               & 2                        & 0.1            \\
75    & 0                    & 5.00                & 5\%               & 2                        & 0.1            \\
76    & 192                  & 5.00                & 5\%               & 2                        & 0              \\
77    & 182                  & 5.00                & 5\%               & 2                        & 0              \\
78    & 146                  & 5.00                & 5\%               & 2                        & 0              \\
79    & 133                  & 5.00                & 5\%               & 2                        & 0              \\
80    & 131                  & 5.00                & 5\%               & 2                        & 0              \\
81    & 130                  & 5.00                & 5\%               & 2                        & 0              \\
82    & 124                  & 5.00                & 5\%               & 2                        & 0              \\
83    & 106                  & 5.00                & 5\%               & 2                        & 0              \\
84    & 101                  & 5.00                & 5\%               & 2                        & 0              \\
85    & 95                   & 5.00                & 5\%               & 2                        & 0              \\
86    & 85                   & 5.00                & 5\%               & 2                        & 0              \\
87    & 81                   & 5.00                & 5\%               & 2                        & 0              \\
88    & 72                   & 5.00                & 5\%               & 2                        & 0              \\
89    & 70                   & 5.00                & 5\%               & 2                        & 0              \\
90    & 61.5                 & 5.00                & 5\%               & 2                        & 0              \\
91    & 60                   & 5.00                & 5\%               & 2                        & 0              \\
92    & 57.8                 & 5.00                & 5\%               & 2                        & 0              \\
93    & 55                   & 5.00                & 5\%               & 2                        & 0              \\
94    & 50.7                 & 5.00                & 5\%               & 2                        & 0              \\
95    & 47.3                 & 5.00                & 5\%               & 2                        & 0              \\
96    & 45                   & 5.00                & 5\%               & 2                        & 0              \\
97    & 44.8                 & 5.00                & 5\%               & 2                        & 0              \\
98    & 42.1                 & 5.00                & 5\%               & 2                        & 0              \\
99    & 42                   & 5.00                & 5\%               & 2                        & 0              \\
100   & 38.5                 & 5.00                & 5\%               & 2                        & 0              \\
101   & 38                   & 5.00                & 5\%               & 2                        & 0              \\
102   & 37.9                 & 5.00                & 5\%               & 2                        & 0              \\
103   & 35                   & 5.00                & 5\%               & 2                        & 0              \\
104   & 32.5                 & 5.00                & 5\%               & 2                        & 0              \\
105   & 32                   & 5.00                & 5\%               & 2                        & 0              \\
106   & 30                   & 5.00                & 5\%               & 2                        & 0              \\
107   & 30                   & 5.00                & 5\%               & 2                        & 0              \\
108   & 27                   & 5.00                & 5\%               & 2                        & 0              \\
109   & 26                   & 5.00                & 5\%               & 2                        & 0              \\
110   & 25                   & 5.00                & 5\%               & 2                        & 0              \\
111   & 25                   & 5.00                & 5\%               & 2                        & 0              \\
112   & 25                   & 5.00                & 5\%               & 2                        & 0              \\
113   & 24                   & 5.00                & 5\%               & 2                        & 0              \\
114   & 24                   & 5.00                & 5\%               & 2                        & 0              \\
115   & 21.2                 & 5.00                & 5\%               & 2                        & 0              \\
116   & 21                   & 5.00                & 5\%               & 2                        & 0              \\
117   & 21                   & 5.00                & 5\%               & 2                        & 0              \\
118   & 20                   & 5.00                & 5\%               & 2                        & 0              \\
119   & 19.7                 & 5.00                & 5\%               & 2                        & 0              \\
120   & 19.1                 & 5.00                & 5\%               & 2                        & 0              \\
121   & 17                   & 5.00                & 5\%               & 2                        & 0              \\
122   & 16                   & 5.00                & 5\%               & 2                        & 0              \\
123   & 15                   & 5.00                & 5\%               & 2                        & 0              \\
124   & 15                   & 5.00                & 5\%               & 2                        & 0              \\
125   & 14                   & 5.00                & 5\%               & 2                        & 0              \\
126   & 14                   & 5.00                & 5\%               & 2                        & 0              \\
127   & 13.5                 & 5.00                & 5\%               & 2                        & 0              \\
128   & 12.9                 & 5.00                & 5\%               & 2                        & 0              \\
129   & 12                   & 5.00                & 5\%               & 2                        & 0              \\
130   & 11.5                 & 5.00                & 5\%               & 2                        & 0              \\
131   & 11                   & 5.00                & 5\%               & 2                        & 0              \\
132   & 11                   & 5.00                & 5\%               & 2                        & 0              \\
133   & 9.8                  & 5.00                & 5\%               & 2                        & 0              \\
134   & 9.5                  & 5.00                & 5\%               & 2                        & 0              \\
135   & 8.9                  & 5.00                & 5\%               & 2                        & 0              \\
136   & 8.6                  & 5.00                & 5\%               & 2                        & 0              \\
137   & 8.5                  & 5.00                & 5\%               & 2                        & 0              \\
138   & 8.5                  & 5.00                & 5\%               & 2                        & 0              \\
139   & 7.8                  & 5.00                & 5\%               & 2                        & 0              \\
140   & 7.6                  & 5.00                & 5\%               & 2                        & 0              \\
141   & 6.9                  & 5.00                & 5\%               & 2                        & 0              \\
142   & 6.8                  & 5.00                & 5\%               & 2                        & 0              \\
143   & 6.8                  & 5.00                & 5\%               & 2                        & 0              \\
144   & 6.7                  & 5.00                & 5\%               & 2                        & 0              \\
145   & 6.5                  & 5.00                & 5\%               & 2                        & 0              \\
146   & 6.5                  & 5.00                & 5\%               & 2                        & 0              \\
147   & 6.3                  & 5.00                & 5\%               & 2                        & 0              \\
148   & 6.3                  & 5.00                & 5\%               & 2                        & 0              \\
149   & 6                    & 5.00                & 5\%               & 2                        & 0              \\
150   & 5.6                  & 5.00                & 5\%               & 2                        & 0              \\
151   & 5.5                  & 5.00                & 5\%               & 2                        & 0              \\
152   & 5.4                  & 5.00                & 5\%               & 2                        & 0              \\
153   & 4.6                  & 5.00                & 5\%               & 2                        & 0              \\
154   & 4.6                  & 5.00                & 5\%               & 2                        & 0              \\
155   & 4.5                  & 5.00                & 5\%               & 2                        & 0              \\
156   & 4.5                  & 5.00                & 5\%               & 2                        & 0              \\
157   & 4.5                  & 5.00                & 5\%               & 2                        & 0              \\
158   & 4.19                 & 5.00                & 5\%               & 2                        & 0              \\
159   & 3.8                  & 5.00                & 5\%               & 2                        & 0              \\
160   & 3.7                  & 5.00                & 5\%               & 2                        & 0              \\
161   & 3.7                  & 5.00                & 5\%               & 2                        & 0              \\
162   & 3.6                  & 5.00                & 5\%               & 2                        & 0              \\
163   & 3.3                  & 5.00                & 5\%               & 2                        & 0              \\
164   & 3.2                  & 5.00                & 5\%               & 2                        & 0              \\
165   & 3.2                  & 5.00                & 5\%               & 2                        & 0              \\
166   & 3.01                 & 5.00                & 5\%               & 2                        & 0              \\
167   & 3                    & 5.00                & 5\%               & 2                        & 0              \\
168   & 3                    & 5.00                & 5\%               & 2                        & 0              \\
169   & 3                    & 5.00                & 5\%               & 2                        & 0              \\
170   & 3                    & 5.00                & 5\%               & 2                        & 0              \\
171   & 2.9                  & 5.00                & 5\%               & 2                        & 0              \\
172   & 2.8                  & 5.00                & 5\%               & 2                        & 0              \\
173   & 2.5                  & 5.00                & 5\%               & 2                        & 0              \\
174   & 2.4                  & 5.00                & 5\%               & 2                        & 0              \\
175   & 2.3                  & 5.00                & 5\%               & 2                        & 0              \\
176   & 2.1                  & 5.00                & 5\%               & 2                        & 0              \\
177   & 2                    & 5.00                & 5\%               & 2                        & 0              \\
178   & 2                    & 5.00                & 5\%               & 2                        & 0              \\
179   & 1.9                  & 5.00                & 5\%               & 2                        & 0              \\
180   & 1.8                  & 5.00                & 5\%               & 2                        & 0              \\
181   & 1.8                  & 5.00                & 5\%               & 2                        & 0              \\
182   & 1.8                  & 5.00                & 5\%               & 2                        & 0              \\
183   & 1.8                  & 5.00                & 5\%               & 2                        & 0              \\
184   & 1.8                  & 5.00                & 5\%               & 2                        & 0              \\
185   & 1.8                  & 5.00                & 5\%               & 2                        & 0              \\
186   & 1.8                  & 5.00                & 5\%               & 2                        & 0              \\
187   & 1.8                  & 5.00                & 5\%               & 2                        & 0              \\
188   & 1.8                  & 5.00                & 5\%               & 2                        & 0              \\
189   & 1.6                  & 5.00                & 5\%               & 2                        & 0              \\
190   & 1.6                  & 5.00                & 5\%               & 2                        & 0              \\
191   & 1.5                  & 5.00                & 5\%               & 2                        & 0              \\
192   & 1.5                  & 5.00                & 5\%               & 2                        & 0              \\
193   & 1.5                  & 5.00                & 5\%               & 2                        & 0              \\
194   & 1.42                 & 5.00                & 5\%               & 2                        & 0              \\
195   & 1.4                  & 5.00                & 5\%               & 2                        & 0              \\
196   & 1.3                  & 5.00                & 5\%               & 2                        & 0              \\
197   & 1.3                  & 5.00                & 5\%               & 2                        & 0              \\
198   & 1.2                  & 5.00                & 5\%               & 2                        & 0              \\
199   & 1.2                  & 5.00                & 5\%               & 2                        & 0              \\
200   & 1.2                  & 5.00                & 5\%               & 2                        & 0              \\
201   & 1.2                  & 5.00                & 5\%               & 2                        & 0              \\
202   & 1.2                  & 5.00                & 5\%               & 2                        & 0              \\
203   & 1.2                  & 5.00                & 5\%               & 2                        & 0              \\
204   & 1.1                  & 5.00                & 5\%               & 2                        & 0              \\
205   & 1                    & 5.00                & 5\%               & 2                        & 0              \\
206   & 1                    & 5.00                & 5\%               & 2                        & 0              \\
207   & 1                    & 5.00                & 5\%               & 2                        & 0              \\
208   & 1                    & 5.00                & 5\%               & 2                        & 0              \\
209   & 0.9                  & 5.00                & 5\%               & 2                        & 0              \\
210   & 0.9                  & 5.00                & 5\%               & 2                        & 0              \\
211   & 0.75                 & 5.00                & 5\%               & 2                        & 0              \\
212   & 0.7                  & 5.00                & 5\%               & 2                        & 0              \\
213   & 50                   & 15.59               & 5\%               & 2                        & 0              \\
214   & 880                  & 16.18               & 1\%               & 2                        & 0              \\
215   & 880                  & 16.18               & 1\%               & 2                        & 0              \\
216   & 496                  & 16.18               & 1\%               & 2                        & 0              \\
217   & 496                  & 16.18               & 1\%               & 2                        & 0              \\
218   & 13.2                 & 22.51               & 5\%               & 2                        & 0              \\
219   & 565                  & 25.36               & 5\%               & 2                        & 0              \\
220   & 148                  & 26.07               & 5\%               & 2                        & 0.1            \\
221   & 12.3                 & 26.13               & 5\%               & 2                        & 0.1            \\
222   & 80                   & 31.69               & 5\%               & 2                        & 0.1            \\
223   & 81.4                 & 32.40               & 5\%               & 2                        & 0.1            \\
224   & 163                  & 34.22               & 5\%               & 2                        & 0.1            \\
225   & 218                  & 36.71               & 5\%               & 2                        & 0.1            \\
226   & 210                  & 37.57               & 5\%               & 2                        & 0.1            \\
227   & 75                   & 38.95               & 5\%               & 2                        & 0.1            \\
228   & 89                   & 39.01               & 5\%               & 2                        & 0.1            \\
229   & 16                   & 39.38               & 5\%               & 2                        & 0.1            \\
230   & 86                   & 40.37               & 5\%               & 2                        & 0.1            \\
231   & 86                   & 40.37               & 5\%               & 2                        & 0.1            \\
232   & 61                   & 42.23               & 5\%               & 2                        & 0.1            \\
233   & 154                  & 44.60               & 25\%              & 2                        & 0.1            \\
234   & 190                  & 48.78               & 25\%              & 2                        & 0.1            \\
235   & 112                  & 51.14               & 25\%              & 2                        & 0.1            \\
236   & 108                  & 52.00               & 25\%              & 2                        & 0.1            \\
237   & 470                  & 53.00               & 25\%              & 2                        & 0.1            \\
238   & 15                   & 54.00               & 25\%              & 2                        & 0.1            \\
239   & 234                  & 56.00               & 25\%              & 2                        & 0.1            \\
240   & 37.5                 & 57.00               & 25\%              & 2                        & 0.1            \\
241   & 60                   & 58.00               & 25\%              & 2                        & 0.1            \\
242   & 160                  & 61.01               & 25\%              & 2                        & 0.1            \\
243   & 50                   & 61.03               & 25\%              & 2                        & 0.1            \\
244   & 62.2                 & 61.24               & 25\%              & 2                        & 0.1            \\
245   & 32                   & 62.45               & 25\%              & 2                        & 0.1            \\
246   & 147                  & 62.49               & 25\%              & 2                        & 0.1            \\
247   & 129                  & 63.77               & 25\%              & 2                        & 0.1            \\
248   & 76                   & 63.88               & 25\%              & 2                        & 0.1            \\
249   & 12                   & 64.21               & 25\%              & 2                        & 0.1            \\
250   & 6                    & 65.79               & 25\%              & 2                        & 0.1            \\
251   & 45                   & 65.81               & 25\%              & 2                        & 0.1            \\
252   & 70                   & 67.17               & 25\%              & 2                        & 0.1            \\
253   & 125                  & 67.20               & 25\%              & 2                        & 0.1            \\
254   & 27.7                 & 69.94               & 25\%              & 2                        & 0.1            \\
255   & 10                   & 71.00               & 25\%              & 2                        & 0.1            \\
256   & 22                   & 73.44               & 25\%              & 2                        & 0.1            \\
257   & 49                   & 73.87               & 25\%              & 2                        & 0.1            \\
258   & 5.5                  & 74.78               & 25\%              & 2                        & 0.1            \\
259   & 6.2                  & 76.46               & 25\%              & 2                        & 0.1            \\
260   & 50                   & 76.78               & 25\%              & 2                        & 0.1            \\
261   & 58                   & 78.01               & 25\%              & 2                        & 0.1            \\
262   & 20                   & 79.65               & 25\%              & 2                        & 0.1            \\
263   & 20                   & 81.42               & 25\%              & 2                        & 0.1            \\
264   & 12.5                 & 83.33               & 25\%              & 2                        & 0.1            \\
265   & 20                   & 84.54               & 25\%              & 2                        & 0.1            \\
266   & 15                   & 85.45               & 25\%              & 2                        & 0.1            \\
267   & 4.6                  & 89.42               & 25\%              & 2                        & 0.1            \\
268   & 60                   & 89.57               & 25\%              & 2                        & 0.1            \\
269   & 85                   & 90.40               & 25\%              & 2                        & 0.1            \\
270   & 8.4                  & 91.40               & 25\%              & 2                        & 0.1            \\
271   & 32.4                 & 92.40               & 25\%              & 2                        & 0.1            \\
272   & 6                    & 93.19               & 25\%              & 2                        & 0.1            \\
273   & 4.5                  & 97.05               & 25\%              & 2                        & 0.1            \\
274   & 17.5                 & 98.45               & 25\%              & 2                        & 0.1            \\
275   & 14.42                & 100.78              & 25\%              & 2                        & 0              \\
276   & 11                   & 102.42              & 25\%              & 2                        & 0              \\
277   & 12                   & 115.96              & 25\%              & 2                        & 0              \\
278   & 4.5                  & 121.87              & 25\%              & 2                        & 0              \\
279   & 9                    & 124.92              & 25\%              & 2                        & 0              \\
280   & 1.4                  & 128.62              & 25\%              & 2                        & 0              \\
281   & 14.5                 & 140.00              & 50\%              & 2                        & 0              \\
282   & 3.7                  & 141.00              & 50\%              & 2                        & 0              \\
283   & 4                    & 170.00              & 50\%              & 2                        & 0              \\
284   & 0.6                  & 180.00              & 50\%              & 2                        & 0              \\
285   & 102                  & 190.00              & 50\%              & 1                        & 0              \\
286   & 100.5                & 200.00              & 50\%              & 1                        & 0              \\
287   & 332.6                & 210.00              & 50\%              & 1                        & 0              \\
288   & 180                  & 220.00              & 50\%              & 1                        & 0              \\
289   & 60                   & 230.00              & 50\%              & 1                        & 0              \\
290   & 26                   & 240.00              & 50\%              & 1                        & 0              \\
291   & 52                   & 250.00              & 50\%              & 1                        & 0              \\
292   & 5                    & 255.00              & 50\%              & 2                        & 0              \\
293   & 28                   & 280.00              & 50\%              & 2                        & 0              \\
294   & 4.1                  & 300.00              & 50\%              & 2                        & 0              \\
295   & 4.5                  & 350.00              & 50\%              & 2                        & 0              \\
296   & 15                   & 400.00              & 50\%              & 2                        & 0              \\
297   & 4.1                  & 450.00              & 50\%              & 2                        & 0              \\
298   & 4.5                  & 500.00              & 50\%              & 2                        & 0              \\
299   & 50                   & 500.00              & 50\%              & 1                        & 0              \\
300   & 118                  & 500.00              & 50\%              & 1                        & 0              \\
301   & 40                   & 500.00              & 50\%              & 1                        & 0              \\
302   & 40                   & 500.00              & 50\%              & 1                        & 0              \\
303   & 40                   & 500.00              & 50\%              & 1                        & 0              \\
304   & 27                   & 500.00              & 50\%              & 1                        & 0              \\
305   & 1000                 & 1000                & 100\%             & 1                        & 0              \\
306   & 2000                 & 3000                & 100\%             & 1                        & 0           
\label{sys_config}
\end{longtable}

\FloatBarrier










\bibliographystyle{elsarticle-num}

\vfill


\end{document}